\documentclass[12pt,preprint]{aastex}

\slugcomment{Draft: \today}

\shorttitle{New nearby stars in the Galactic Plane}
\shortauthors{Reid et al}

\begin{document}

\def\pedant{Mart{\'{\i}}n }
\def\etal{{\sl et al.}}
\def\etall{{\sl et al. }}
\def\pma{$\arcsec$~yr$^{-1}$ }
\def\kms{km~s$^{-1}$ }
\def\msun{$M_{\odot}$}
\def\rsun{$R_{\odot}$}
\def\lsun{$L_{\odot}$}
\def\halpha{H$\alpha$}
\def\hbeta{H$\beta$}
\def\hgama{H$\gamma$}
\def\hdelta{H$\delta$}
\def\Teff{T$_{eff}$}
\def\logg{$log_g$} 

\title {Meeting the Cool Neighbours, VI:
A search for nearby ultracool dwarfs in the Galactic Plane}

\author{I. Neill Reid}
\affil{Space Telescope Science Institute, 3700 San Martin Drive, Baltimore,
MD 21218 \\
and \\
Department of Physics and Astronomy, University of Pennsylvania, 209 South 
33rd Street, Philadelphia, PA 19104}
\email{inr@stsci.edu}

\begin {abstract}

Surveys for nearby low-luminosity dwarfs tend to avoid the crowded
regions of the Galactic Plane.
We have devised near-infrared colour-magnitude and colour-colour selection criteria
designed to identify late-type M and early-type L dwarfs within
12 parsecs of the Sun. We use those criteria to search
for candidates within the regions of the Galactic Plane ($|b| < 10^o$) covered
by the Second Incremental Release of data from the Two-Micron All Sky Survey.
Detailed inspection of the available photographic images of the resulting
1299 candidates confirms only two as ultracool dwarfs. Both are known
proper motion stars, identified in the recent survey 
by L\'epine \etal (2002). Despite the low numbers, the inferred surface
density is consistent with comparable surveys at higher latitudes. We discuss
the implications for the luminosity function, and consider means of improving the
efficiency and scope of photometric surveys in the Plane.

\end{abstract}

\keywords{stars: low-mass, brown dwarfs; stars: luminosity function, mass function; 
 Galaxy: stellar content }

\section {Introduction}

Most of the stellar systems known to lie within the immediate Solar Neighbourhood
were identified originally in proper motion surveys, notably Willem
Luyten's analyses of photographic plates taken with Palomar 48-inch Schmidt 
(Luyten, 1979; 1980, and references 
therein). The high star densities present near the Galactic Plane have long been 
recognised as a hindrance to that type of survey, particularly in searches for stars
of intrinsically low luminosity. Thus, while analysis of current nearby-star catalogues,
such as the preliminary Third Nearby Star Catalogue (Gliese \& Jahreiss, 1991: pCNS3),
shows no evidence for bias at low
Galactic latitudes in the distribution of early- and mid-type M dwarfs, there is a 
clear deficit of later-type (spectral types $\ge$ M5) dwarfs (Reid \etal, 2002a: PMSU4). 
This deficiency has been highlighted by L\'epine \etal's (2002) recent proper
motion survey of the northern Galactic Plane, which uses image subtraction
to compare scans of first- and second-epoch sky survey plates. 
Of the 601 stars identified in that survey
with $\mu > 0.5$\pma, only 460 are included in the Luyten Half Second (LHS)
catalogue (Luyten, 1980), even though all are 
visible on plate material taken for the first Palomar Sky Survey (POSS I).

Photometric surveys provide a complementary and, in some cases,
more effective means of surveying the Galactic Plane for nearby, late-type dwarfs. 
Proper motion surveys require detection at several epochs, and searches for faint
red stars are often limited by the availability of only blue-sensitive imaging data at
early epochs. Photometric surveys can be tuned to the longer wavelengths where 
low-mass dwarfs are most luminous, and can avoid the (often exagerated)
potential for kinematic bias inherent in proper motion selection.
In practice, most surveys use both techniqes, either using photometry or
spectroscopy to confirm proper-motion selected candidates or, as described
further below, using proper motion to identify nearby dwarfs in a
photometrically-selected sample.

To be most effective, photometric surveys should span
a wide wavelength range, with near-contemporary observations in the various passbands. 
The latter attribute is
necessary to allow reliable matching of separate observations of stars with sensible proper motions;
the former is required since other sources, such as distant giants, reddened stars and young T Tauri stars,
can mimic M dwarf colours in some passbands, and those contaminants greatly outnumber genuine
nearby stars at low latitudes. Unfortunately, at present we lack the full complement of
such surveys. The Two-Micron All-Sky Survey (2MASS: Skrutskie \etal, 1997) provides high quality
astrometry and photometry at near-infrared wavelengths, and far-red I-band data are now available 
for almost the full sky, through the second generation UK Schmidt and POSS II surveys (Reid \etal, 1991).
Scans of the latter plates can provide astrometrically accurate data for well-resolved sources, 
with photometric uncertainties of $\pm0.25$ magnitudes to I$\sim18.5$ magnitudes. However,
published optical catalogues based on Schmidt surveys  (e.g. USNO-A2 catalogue, Monet
\etal, 1998) generally require detection in two or more passbands. As a 
result, M-dwarf surveys are often limited
by the sensitivity of the blue-passband 103aO or IIIaJ survey plates. 

Given these technical constraints, our initial survey for previously-undetected ultracool
dwarfs in the immediate Solar Neighbourhood (d $< 20$ pc) was 
limited to $|b| > 10^o$, where infrared data alone suffice (Cruz \etal, 2003; Paper V). 
This paper outlines our first attempt to extend that survey to lower latitudes, albeit for
a more restricted range of distance and spectral type. Section 2 outlines our
selection criteria; section 3 describes our results; section 4 discusses the nature of the
faint proper motion stars identified recently by L\'epine \etal (2002); and section 5 presents our conclusions.

\section {Selection Criteria }

We use 2MASS near-infrared photometry as the basis for the initial selection
of candidate ultracool dwarfs (spectral types later than M7). The 2MASS Second Incremental Data Release  
(hereinafter referred to as the 2M2nd) covers 48\% of the sky, including $\sim7$\% 
within 10 degrees of the Galactic Plane. As discussed in Paper V, and illustrated in
Figure 1, (J-K$_S$) colours provide a means of identifying late-type M and L dwarfs, and
deriving moderately-accurate photometric parallaxes.  
At high galactic latitude, we used J magnitudes and JHK colours to isolate $\sim2000$ objects
with photometric properties
consistent with dwarfs within 20-parsecs of the Sun (the 2MU2 sample). Unfortunately, those
criteria fail at lower latitudes: nearly 1 million sources with $|b| < 10^o$ 
meet the colour/magnitude criteria outlined in Paper V. Clearly, it is not possible
to carry out the same kind of detailed follow-up observations for such a large sample, so
we have adopted different tactics in our low-latitude search.

\subsection {Colour-magnitude and colour-colour selection}

As a first cut, we set colour limits of $1.0 < (J-K_S) < 2.1$. The blue limit
corresponds to a spectral type of $\approx$M7, while the red limit matches
the reddest L dwarfs. We fit a linear relation in the (M$_J$, (J-K$_S$)) plane
(Figure 1), 
\begin{displaymath}
M_J \ = \ 4.2 \ (J-K_S) \ + \ 6.6, \ \sigma = 0.28 \ {\rm mag.}
\end{displaymath}
and use that to eliminate source with photometric parallaxes, $\pi_{J-K}$,
exceeding 83 milliarcseconds (i.e. $d_{J-K} > 12$ parsecs). At the same time,
we reject sources where the 2MASS data indicate possible contamination by
diffraction spikes, diffuse emission or image artefacts (i.e. we require the 2MASS
parameter $cc\_flg = 000$). We have also eliminated sources where one or 
more of the formal uncertainties in the measured magnitudes is set to `null',
on the basis that those are likely to haver unreliable photometry.

Our use of (J, (J-K$_S$)) selection criteria means that we are dealing with a 
magnitude limited sample. As a result, Malmquist bias is introduced.
The distances involved are sufficiently small that we can assume a 
uniform density distribution, so the classical result applies:
\begin{displaymath}
\bar M \ = \ M_0 \ - \ 1.38 \sigma^2 \ = \ M_0 - 0.11 
\end{displaymath}
where $M_0$ is the true absolute magnitude for a given (J-K$_S$) colour, and $\bar 
M$ the actual mean absolute magnitude in an empirically-selected sample. Near 
our chosen distance limit, stars with $M_J > M_0$ scatter out of the sample, 
while stars with $M_J < M_0$ and $d > 12$ parsecs scatter into the sample. The 
net effect is relatively minor: the 0.11 magnitude offset in $\bar M$ 
corresponds to an increase in the effective distance limit of only $\sim5\%$, or 
$\sim16\%$ in the effective sampling volume. Moreover, the colour-magnitude
relation proves to serve more as a selection criterion than a distance estimator.

Finally, we include the (J-H)/(H-K$_S$) colour-colour selection criteria
used to select the 2MU2 sample (Paper V). Those are designed to minimise
giant contamination, while spanning the full L dwarf sequence.
The resultant sample of low-latitude candidates includes $\sim170,000$ sources
 -  still too many for detailed follow-up.

\subsection {Optical counterparts in the 2MASS catalogue}

The 2MASS point-source catalogue includes optical data, primarily from the USNO-A2 catalogue
(Monet \etal, 1998). The latter catalogue was constructed by cross-referencing scans
of the POSS I blue (103aO) and red (103aE) plates with field centres
north of  declination $ -18^o$,
and the UKST blue (IIIaJ) and ESO red (IIIaF) plates centred at $\delta \le -20^o$.
Sources are included in USNOA2.0 only if they appear on both blue and red plates
(positions coincide to better than 2 arcseconds). 

A match between a 2MASS source and an object in the USNO-A2.0 catalogue 
requires that the positions agree to better than 
5 arcseconds\footnote{USNO was responsible for defining the 2MASS co-ordinate system, so the
two catalogues have consistent astrometry.}.  The average epoch of the 2MASS data is 1998, while
the mean epochs are $\sim1954$ for POSS I and $\sim1982$ for UKST/ESO plate material.
Thus, the requirement for positional coincidence translates to a requirement that the
proper motion, $\mu$, is less than $\sim0.11$\pma for $\delta > -18^o$
and $\mu < 0.28$\pma for more southern sources. At 12 parsecs, those motions correspond to
tangential velocities $V_{tan} < 6.25$\kms and $<16$\kms, respectively. Based on the analysis in
PMSU4, fewer than 3\% of local disk dwarfs are expected to meet the former criterion, while
only 15\% meet the latter. Thus, in our search for new nearby stars, 
we have added the requirement that candidates
have no USNO-A2.0 optical counterpart. This reduces the target list to 70,362 sources.

\subsection {IRAS sources}

Since the USNO-A2 catalogue requires detection on both red and blue plates, 
highly-reddened late-type giants and protostars in the 2MASS database are likely to lack optical counterparts.
Even though many of these objects are visible on the POSS I and ESO red plates, the enshrouding
circumstellar dust leads to blue magnitudes below the limit of the 103aO or IIIaJ plate material.
We can eliminate the brightest of those sources by searching for counterparts in
the IRAS point source catalogue, since even the nearest late-M and L dwarfs have far-infrared fluxes
well below the IRAS detection limits.

Eight thousand one hundred and thirty of our Galactic Plane nearby-star candidates lie 
within 60
arcseconds of an IRAS source. Figure 2 plots the number of sources as a function of 
positional difference, $\Delta$. There is a strong peak at small separations, with a minimum at
$\sim 13$ arcseconds. We have therefore excluded candidates with $\Delta < 12$ arcseconds
as likely to be associated with the IRAS source. This eliminates only 1288 sources.

\subsection {A candidate list of late-M/early-L dwarfs}

Even after applying the above selection criteria, our list of potential nearby 
ultracool dwarfs still includes
over 69,000 candidates. Cursory inspection of digitised sky survey data for a randomly chosen
subset of these sources shows that many, probably most, are visible on at least the IVN (I-band)
and the IIIaF (R-band) plates taken for the second epoch POSS II and AAO Schmidt surveys, while
some are also visible on the POSS I E plate.
Even with uncertainties of $\pm0.25$ magnitudes in the optical photometry, combining the 2MASS data with 
an R/I point-source catalogue would provide an effective method of eliminating a substantial fraction
of the contaminating evolved giants, reddened stars and young protostars in our Galactic Plane
sample. Unfortunately, such catalogues have not yet been constructed. Given those circumstances,
we have modified the goals of the current investigation.

The onset of H$^-$ absorption and pressure-induced formation of H$_2$ 
in M dwarfs leads to main-sequence stars outlining a
characteristic S-curve in the (J-H)/(H-K$_S$) plane 
(as illustrated in the right-hand panel in Figure 3). Those opacity sources are
much less important in low-pressure red giant atmospheres, and, as a result, those stars 
follow a sequence which runs above the dwarf stars (redder (J-H) colours at
a given (H-K$_S$) (Gingerich, Latham, Linsky \& Kumar, 1966; Lee, 1970; Mould \&
Hyland, 1976).
M-type, S-type and C-type asymptotic giant branch stars extend that sequence
to redder colours, and, as Figure 3 shows, overlap with the late-M/L-dwarf sequence
redward of (H-K$_S) \sim 0.4$ 
magnitudes. The reddening vector (from Koorneef, 1983) is almost parallel to the giant sequence, so
both protostars and distant G and K dwarfs in the Galactic Plane can lie close to the 
giant sequence in the JHK plane. As a result, all of these more luminous stars contaminate
our colour-selected Galactic Plane sample, particularly at the redder colours corresponding to
mid- and late-type L dwarfs. Therefore, for this initial study, we have defined colour
limits which isolate late-type M and early-type L dwarfs:
\begin{eqnarray}
(J-H) \quad&  > &  \quad 1.61 (H-K_S) \ - \ 0.305 \nonumber \\
(J-H) \quad&  \le &  \quad 1.875 (H-K_S), \ (H-K) \ < \ 0.48 \nonumber \\
(J-H) \quad&  \le &  \quad 0.9, \ (H-K_S) \ \ge \ 0.48 \nonumber 
\end{eqnarray}
All sources also have $1.0 < (J-K_S) < 2.1$.

Figure 4 illustrates how these JHK$_S$ selection criteria map onto spectral type and colour-colour
distributions, plotting data for nearby stars and brown dwarfs. The primary constituents in our
survey should be dwarfs with spectral types M8 to $\approx$L1/L1.5, although a few
later-type L dwarfs (notably 2M0036+18, L3.5, Reir \etal, 2000) also fall within this colour range.

Applying these selection criteria reduces the list 
of potential nearby ultracool dwarfs from $\sim69,000$ to only 1311 sources,
including 12 duplicates. The final sample therefore includes 1299
candidates. Figure 3 plots
the (J, (J-K$_S$) and (J-H)/(H-K$_S$) distribution of those sources.
The concentration of candidates towards the locus occupied by red giants (and
reddened main-sequence stars)  is obvious in the latter diagram. 

\section {Results}

\subsection {Photographic data for the photometric sample}

Having reduced the number of candidates to a manageable total, the 
final selection process involves 
examining the digitised sky survey images of each of these candidates, comparing
both first and second epoch data against the 2MASS scans. To do so, we used the
downloading facilities provided by CADC website 
({\sl http://cadcwww.dao.nrc.ca/dss/}). All of the first 
epoch images for sources at $\delta > -2^o$ are from POSS I 103aE (red) plate
material (epochs 1950 to 1958); SERC/UKST IIIaJ (blue) plates 
provide the corresponding
data for the more southern targets. POSS II supplies second-epoch 
IIIaF and IVN 
material at $\delta > 2^o.5$, with the plates taken between 1987 and 1997,
while the southern sources are covered by the SERC/UKST Galactic
Plane I/SR survey. Plates for the latter survey were taken
between 1977 and 1984 (Morgan, 1995), with most of the region discussed here
observed in either 1979 or 1980.

What photometric characteristics are we looking for? 
As shown in Figure 4, ultracool dwarfs with spectral
types between M7 and L2 have $5 < (R-J) < 6$. The passband of the POSS I
103aE plates overlaps with only the shorter-wavelength segment 
of the Cousins R-band, so these colour limits need to be adjusted. We 
have used Luyten's data for nearby stars (Figure 4 from Paper I) to derive the
following relation,
\begin{displaymath}
 (m_r-R) \ = (0.25\pm0.06) \ + \ (0.55\pm0.05) (R-I), 
\quad \sigma=0.41 \ {\rm mag, \ 220 \ stars}
\end{displaymath}
M7 dwarfs have (R-I)$\sim$2.3 mag, and that colour saturates at $\sim2.5$ mag
for later-type dwarfs, so the effective colour limits for the current
survey are $6 < (m_r-J) < 7.5$ (allowing some latitude for intrinsic dispersion).
The limiting magnitude of the POSS I 103aE platescans is m$_r \sim 20.5$, so,
with J$_{max}\sim 13.5$ (Figure 3), 
all save the faintest candidates should be visible
on the first epoch plates. However, with (B-R)$\sim$4 for ultracool dwarfs and
B$_J$(lim)$\sim22.5$, a higher proportion of southern targets are 
likely to be absent from first epoch IIIaJ material.

Considering the second epoch plate material, the ultracool dwarfs targeted here
have J$<$ 13.5 (Figure 3) and (I-J) colours in the range $2.8 < (I-J) < 4$ (Figure 4).
Thus, any such objects within 12 parsecs should be detected readily
on the IVN plate material (I$_{lim} \sim 18.5$). Moreover, almost all
should also be visible on the IIIaF plates (R$_{lim} \sim 21.5$). 

Finally, we do not expect exact positional coincidence between either 
first-epoch or second-epoch photographic images of genuine nearby dwarfs. The
Sun has a velocity of 11 kms$^{-1}$ with respect to the Local Standard of Rest, and the
average tangential velocity of nearby disk dwarfs is $\sim30$ kms$^{-1}$. A tangential
velocity of only 5 kms$^{-1}$ produces an annual proper motion of $\sim0.08''$ yr$^{-1}$
at 12 parsecs. This corresponds to a displacement of $\sim3.5$ arcseconds between
POSS I and the 2MASS imaging and $\sim1.5$ arcseconds 
between the SERC/UKST Galactic Plane survey and the 2MASS data.

In summary, genuine ultracool dwarfs within 12 parsecs are not expected to have
optical counterparts at the exact 2MASS position on either 
the first epoch plate material or the SERC/UKST second-epoch plate material. 
However, those dwarfs are expected to be 
visible on all IVN plates and on most IIIaF and POSS I 103aE plate material.
Therefore, if there is no first-epoch optical candidate at the 2MASS position, we
should be able to identify an optical point-source 
image with no 2MASS counterpart. As discussed below, the overwhelming majority of
candidates fail to meet these expectations.

\subsection {The first epoch plate material}

Most of the 1299 candidates can be rejected based on simple visual inspection 
of the first epoch plate material. Approximately half
of the sample (497 sources) have an isolated, point-like optical
counterpart on the first epoch images. In most cases, those sources
are sufficiently red to be invisible on the blue plate
material, and hence absent from the USNO catalogue. 
In some cases, the image is barely discernible on even the
103aE plate. The extremely red (m$_r$-J) colours of many
of those sources fall outside the range expected for ultracool dwarfs,
and suggest significant foreground (or circumstellar) reddening, while 
the positional coincidence with the 2MASS image argues strongly against
proximity to the Sun.

In 353 cases, visual inspection shows that two (or three) stars of similar brightness, 
separated by $<5$ arcseconds, lie at the location of the 2MASS source.
It is likely that those stars are unresolved in the USNO
scans of the photographic plate material, leading to incorrect
co-ordinates and a failure to match one or more sources against the 2MASS database.
Similarly, 311 sources lie in extremely crowded regions, mainly close to
the Galactic Centre, where the high star density complicates
automated analysis of the optical data. None
of these candidates show evidence for significant relative
motion between the first and/or second epoch DSS images and the 2MASS
data. Finally, 37 sources are embedded in nebulosity on the 103aE, IIIaJ and
IIIaF (but not IVN) plates.

\subsection {The second epoch plate material}

After eliminating all of the above spurious candidates, only 102 sources 
lack obvious optical counterparts on the first epoch plate material.
Figure 5 plots the ($\alpha, \delta$) distribution of the initial
1299 sources, and the reduced sample. Almost all
of the latter lie in the vicinity of the Galactic Centre, with 
only four sources in the northern (equatorial) hemisphere. All ninety-eight 
of the southern sources are covered by the UKST/SERC Galactic Plane
survey, and all have R/I counterparts within 
1 arcseconds of the 2MASS location on both IIIaF and IVN second-epoch
plate material. As noted above, this indicates a low proper motion, 
$\mu < 0.06$\pma, corresponding to V$_{tan} < 3$ kms$^{-1}$ for a distance of 12 parsecs. 
In a few cases, a very faint image is discernible on the first epoch plates
(see comments in table 1); in others, the IIIaF image is 
substantially brighter than expected, given the absence of a clear detection on
the first-epoch plate material. The last objects are probably red giant variables, 
which happened to have been at minimum during the initial survey observation. 
The absence of significant proper motion for any of these southern sources
strongly suggests that all 98 are distant giants. 

Of the four northern candidates, two can be ruled out as probable nearby star
candidates. 2M1941443+220904 has a counterpart on the POSS II IIIaF and IVN plates,
while 2M2009142+331448 is clearly visible on both POSS I and POSS II plate material, although
the POSS I image is affected by a plate detect. 
The remaining two sources meet all our selection criteria and are confirmed as
nearby dwarfs. Both have an obvious optical counterpart on 
the first and second epoch plate material, offset from the 2MASS source.
Both targets, 2MASSI 0539248+403844 
and 2MASSI 0602305+391059, have already been identified as nearby dwarfs
through the Galactic Plane proper motion survey undertaken by L\'epine \etal\ (2002). L\'epine \etal\ (2003) estimate a spectral type of M8 for 2M0539.
As part of our NStars follow-up program (described in 
Paper V), we obtained spectroscopy of that star (Figure 6) and estimate
a spectral type of M8.5. Comparing the 
near-infrared photometry against data for VB 10 (M8, M$_K$=9.99) suggests that this
star lies close to 10 parsecs distant from the Sun. 2M0602 is an early-type
L dwarf at a distance of $10.6\pm0.8$ parsecs (Salim \etal, 2003). We also
obtained spectroscopy of LSR0602 (see Salim \etal) and estimate
a spectral type of L1.5.

\section {A comparison with other surveys}

Our survey of the section of the Galactic Plane covered by the 2MASS
Second Incremental Release has netted only two ultracool dwarfs with photometric
properties matching our selection criteria.
To determine whether this meagre detection rate
is reasonable, we need to compare our results against data from other
surveys. We consider three comparisons: first, our own 2MASS-based survey for
ultracool dwarfs at higher galactic latitudes (Paper V); second, Salim \&
Gould's (2003) enhanced version of Luyten's NLTT catalogue; and, finally,
L\'epine \etal's (2002) new proper motion survey of the northern Galactic Plane.

\subsection {High latitude ultracool dwarfs}

Paper V describes our survey for late-type M and L 
dwarfs within 20 parsecs, analysing 2M2nd data covering  $|b| > 10^o$.
The colour/magnitude selection criteria used in this 
high-latitude NStars survey are broader than 
those used in the current survey, and only 202 of the 1672 ultracool candidates 
in the 2MU2 sample fall
within the JHK$_S$ limits outlined in section 2.  
We have either literature identifications or 
observations of all these sources, and the majority are red giants, predominantly drawn 
from the asymptotic giant branch. However, 36 objects are confirmed as
late-type dwarfs, including one M4, six M5s, seven M6s, nine M7s, seven M8s, five M9s and
two L dwarfs (2M0746, L0.5, and 2M0523, L2.5). 
These dwarfs are drawn from 37\%
of the celestial sphere, so we predict an overall surface density of 
$6.9\pm1.1$  late-type dwarfs within the area covered by 
our Galactic Plane survey. Fourteen of the 36 are ultracool dwarfs,
with spectral types M8 or later, so we expect a surface density of 
$2.6\pm0.7$ within our survey. In fact, 
we have succeeded in identifying one M8 and one L1.5.
Thus, our results are in broad agreement with the expected numbers of
M8+ ultracool dwarfs, but it seems likely that $\sim5$ nearby ($d<12$ pc.)
M5 to M7 dwarfs remain to be discovered. 
 
\subsection {The Luyten surveys}

Luyten's proper motion surveys provide an alternative means of
identifying nearby stars. 
One of the main criticisms of the NLTT and LHS 
catalogues is the existence of substantial errors in the astrometry
of a small subset of the proper motion stars.
This is not really surprising, given the scope of Luyten's work and the
many opportunities for transcription errors in the contemporary 
publishing process, and Salim \& Gould (2003) point out that the
overwhelming majority of stars have positions accurate to a few arcseconds.
Nonetheles, the existence of these sporadic outliers complicates any 
statistical analysis. Fortunately, 
with the availability of multi-epoch digitised sky surveys, 
the offending stars can be re-discovered (or, in
some cases, eliminated as spurious) and improved positions provided for
the full catalogue. Thus, Bakos, Sahu, \& N{\' e}meth (2002) have used the DSS to
update the LHS catalogue, while Salim \& Gould have combined
data from the Hipparcos, Tycho and 2MASS catalogues to similarly
enhance the NLTT. We have taken the latter catalogue (designated as SGNLTT)
as our primary reference here.

Salim \& Gould's revised NLTT includes only systems with improved 
astrometry, a total of 36020 of the 58845 stars listed by Luyten. Some 28379 of
those stars have 2MASS K$_S$-band photometry. Most of the remaining stars
lie outwith the region covered by the 2M2nd, 
although $\sim200$ are sufficiently bright that the
2MASS data are saturated. Thirty-one stars have near-infrared
colours and magnitudes which meet the (J, (J-K$_S$)) selection criteria outlined 
in \S2.4. As pointed out by L\'epine \etal, four well-known 
late-type dwarfs, LHS 2397a, 325a, 534a and 3989a, are listed in the LHS
but not the NLTT. LHS 2397a is the only star of these four which falls 
within the region covered by the second incremental release and, at
14 parsecs, it falls below our (J, (J-K$_S$)) selection criterion.  
Data for the SGNLTT stars are listed in Table 2 and plotted in Figure 7.

Taken at face value, Table 2 suggests that our high latitude 2MASS survey 
underestimates the surface density of ultracool dwarfs, particularly since seven of the 31 
SGNLTT stars lie at $|b| < 10^o$ (the latter stars are plotted as solid points in Figure 7). 
However, many of the sources have JHK$_S$ colours inconsistent with our selection 
criteria, and more detailed inspection shows that most of those stars are not 
ultracool dwarfs. 

First, eleven of the 31 dwarfs, including 4 of the 7 stars in the Plane, are 
close doubles (see notes to Table 2), and the 2MASS photometry of these sources is not 
reliable. Optical data indicate that all are more distant, earlier-type M dwarfs. Second, 
visual inspection shows that the remaining three low latitude stars and LP 826-6 are 
incorrectly identified in the SGNLTT. In each case, the designated source shows no 
evidence for significant motion on
the first and second epoch photographic images and the 2MASS scans; 
nor is there an obvious alternative candidate within 5 arcminutes. Given the 
JHK$_S$ colours, we identify these four SGNLTT stars as background giants, unrelated 
to the NLTT proper motion stars.

Five of the remaining sixteen stars are known mid-type M dwarfs: GJ 2005, 
GJ 3590, G180-11, GJ 3981 and Gl 866. Three of these stars meet our JHK$_S$
two-colour criteria. With spectral types between M4 and M5.5, 
the extremely red (J-K$_S$) colours measured for  these relatively bright stars
are somewhat surprising. Further investigation, particularly comparative
near-infrared spectroscopy, would be interesting. 

Thus, twenty of the 31 sources listed in Table 2 are not nearby ultracool dwarfs.
Ten of the remaining eleven stars are included in
the 2MASS photometric sample discussed in Paper V, and are confirmed as late-type 
dwarfs\footnote{ We note that an additional 
two NLTT stars, LP 944-20 and LP 655-48, are identified as ultracool dwarfs
in the 2MASS survey. LP 944-20 lies south of the $\delta=-33^o$ limit of
SGNLTT; LP 655-48 is included in the SGNLTT, but an error in Luyten's
proper motion precluded a match with the 2MASS data}. 
These include the M9.5 dwarf BRI0021-0214, 
for which 2MASS measures (J-H)= 0.94, leading to the star failing the current
selection criteria. In fact, the J-band 2MASS survey image of this dwarf is
affected by a meteor, and the true (J-H) colour is $\sim0.72$ magnitudes (Irwin
\etal, 1991). 
The final star is LHS 6234, which has (m$_r$-J)$\sim 3.9$, suggesting that
it is also a mid-type M dwarf with relatively red near-infrared colours.

In summary, using the NLTT catalogue to search for nearby late-type dwarfs
recovers a subset of the ultracool dwarfs identified in our 2MASS NStars survey,
together with a few more distant mid-type M dwarfs with unusually red 
near-infrared colours and one additional ultracool dwarf, BRI0021. 

\subsection {The LSR sample}

Finally, we compare our results against L\'epine \etal's (2002; LSR)
new survey for high proper motion stars in the northern Milky Way. Taking
advantage of technological advances, LSR have developed a method of 
aligning and differencing images from the digitised versions of the first
and second epoch photographic sky surveys. Subtracting low-motion background
stars greatly reduces the effective crowding, and makes it possible to use
automated blink techniques to identify high-$\mu$ stars. Based on this method,
LSR have identified 601 stars with $0.5 < \mu < 2.0$\pma, 
$9 < R < 20$, $|b| < 25^o$ and
$\delta > 0^o$. Four hundred and sixty of those stars are catalogued by Luyten
in either the LHS or NLTT, but LSR identify 141 stars as new discoveries.
In fact, two of the latter stars are rediscoveries: 
LSR0522+38 is LP 251-35 (not included in either the LHS or NLTT); and 
and LSR0346+25 is WD0346+246, the archetypical cool halo white dwarf
(Hambly \etal, 1997). L\'epine, Rich \& Shara (2003) have recently presented
supplementary spectroscopy of a further 104 stars from this sample.
Since this survey covers a significant fraction of the northern Milky Way, 
it provides an
independent means of testing the completeness of our own photometric
analysis.

Sixty-one of the 141 new proper motion stars in the LSR survey lie
within the area covered by the 2M2nd,
although six (all white dwarfs) are not visible on any of
the 2MASS images. Of these 55 detected stars, thirty-three have $|b| < 10^o$
and overlap with the current survey.
Table 3 lists the 2MASS JHK$_S$ photometry (and upper limits) for
all 61 stars.

As with the Luyten stars, we can use the optical/IR photmetry to determine whether 
any of the LSR stars are ultracool dwarfs which should have been
discovered in our survey. We can also combine the 2MASS near-infrared
data with the optical photometry and astrometry measured by LSR to
determine the likely nature of all of the stars in the sample, supplementing the
available spectroscopy.

Figure 8 plots (J, (J-K$_S$) and (J-H)/(H-K$_S$) diagrams for the LSR sample.
As before, we include data for a variety
of known stellar objects to provide a reference. Inspection of
these diagrams reveals several stars meeting one or other of the 
search criteria we have used to define our ultracool candidates.
However, LSR0602 and LSR0539 are the only
sources which both meet all our photometric criteria and
lie within 10 degrees of the Galactic Plane.

Four other stars have (J-K$_S) > 1.0$. In two cases, the J magnitude is
fainter than our selection limit, and the optical/near-infrared colours
are much bluer than expected for late-type M dwarfs. These are the cool
white dwarf, LSR0346+24; and LSR2105+2514, which Lowrance \etal (2003) 
have shown to be a carbon dwarf. The other two stars have more 
conventional colours. 
LSR1835+3259 has been confirmed spectroscopically as an M8.5 dwarf at a 
distance of less than 6 parsecs (Reid \etal, 2003; L\'epine \etal, 2003), 
but it lies at $b=19^o.2$.

The fourth star, LSR0510+2713\footnote{ The position listed for this
star by LSR is incorrect; we measure approximate co-ordinates of
$\alpha = 05:10:20.1$ $\delta = +27:14:02$ (J2000) from the DSS scans of
the  POSS II IVN plate.}, is an M8 dwarf (Figure 6). 
However, it lies 0.02 magnitudes redder than the (J-H) upper limit of
our colour selection. Increasing our selection limits to include LSR0510+2713 would add a
further 1800 sources to our candidate list, more than doubling the sample size.
Our spectrum of LSR0510+2713 shows emission at H$\alpha$ (EW=44.5 \AA) and the near-infrared
calcium triplet, and the star is coincident with the X-ray source 1RXS J051019.3+271422.
Comparing the photometric properties against those of VB10 suggests a distance of 
$\sim8.2\pm1$ parsecs. 

Approximately two-thirds of the 55 LSR stars have extant spectroscopy (Table 3).
We can determine the nature of the remaining stars by combining the 2MASS 
near-infrared photometry with the observed distribution in the
(R-J)/(J-K$_S$) two-colour diagram and the (H$_R$, (R-J))
reduced proper motion diagram (RPMD: Figure 9). 
Reduced proper motion is defined as 
\begin{displaymath}
H \ = \ m + 5 + 5 \log{\mu} \ = M \ + \ 5 \log{V_T} \ + const. 
\end{displaymath}
where $V_T$ is the tangential velocity (Luyten, 1923). $H$ can
therefore serve as a surrogate for absolute magnitude, and the
RPMD as an HR diagram with velocity-dependent
offsets between different stellar populations. 

In their analysis of the full proper motion sample, LSR considered the
distribution in the (H$_R$, (B-R)) RPMD. This provides adequate
separation between hydrogen-burning stars and white dwarfs (similar 
colours, but different absolute magnitudes), but is less successful
in segregating late-type disk and halo subdwarfs. The disk main-sequence
steepens sharply at spectral types later than $\sim$M2, and intersects
the shallower halo subdwarf sequence at (B-R)$\sim2$ (Gizis, 1997). This
leads to overlapping sequences in the RPMD. The problem can be circumvented
by using more appropriate colours, such as (V-I), where the 
two sequences both maintain a shallower slope and greater separation
over a wider range of spectral types (Reid, 1997). Optical/near-infrared
colours, such as (R-J), are also well suited to this type of population
analysis, as highlighted by Salim \& Gould (2002).

The distribution of the LSR stars in the (H$_R$, (R-J)) and (R-J)/(J-K$_S$)
diagrams are shown in Figure 9. These stars have $\mu > 0.5$\pma, and,
as a reference, we plot data for NLTT stars with similar motions (from
Salim \& Gould, 2003). The two-colour diagram shows that in
most cases the optical/IR colours are broadly consistent with main-sequence
dwarfs (the carbon dwarf, LSR2105+2514, is an obvious outlier).
Three broad sequences are evident in the RPMD
marking the locations of disk white dwarfs, disk main-sequence 
stars and, between the two disk sequences, halo subdwarfs.  As
an additional guide, we plot reduced proper motion
data for known intermediate-abundance
($\langle [Fe/H] \rangle \sim -1$) and extremely metal-poor 
([Fe/H] $>  -1.5$) subdwarfs (from Gizis, 1997 and Reid \etal, 2001).
Two subdwarfs, LHS 2497 and 3409 (both intermediate-abundance sdM dwarfs),
lie amongst the disk dwarfs. 

Based on these reference sequences, we can estimate population 
types for the LSR proper motion stars. Those types are listed in Table 3, 
together with spectroscopic data, where available. By and large, 
the agreement is reasonable, with LSR2010+2938 (H$_R$=16.85, (R-J)=0.73) the only
major discrepancy: we identify it as a possible white dwarf, but spectroscopy
shows it to be an early-type intermediate subdwarf. Seven LSR stars
lie in the vicinity of LHS 3409 in the RPMD, and we identify those stars as
possible sdM subdwarfs (labelled disk/sdM in Table 3). Spectroscopy shows that
all seven are, in fact, intermediate subdwarfs.

For present purposes, 
the most significant aspect of Table 3 is that only 21 of the 57 LSR stars
are classed as disk dwarfs. Most of those stars have colours consistent with 
spectral types between M3 and M6, and, therefore, M$_K < 9$ and distances 
exceeding 12 parsecs. 

\subsection {Results}

Summarising the results from these comparisons, it is clear that
the colour-magnitude and colour-colour limits we have adopted do not
encompass the full range of properties spanned by M8 to L1.5 ultracool dwarfs.
Both BRI0021-0214 (one of ten nearby NLTT dwarfs) and LSR0510+2713 
(one of three LSR ultracool dwarfs) lie outwith our JHK$_S$ colour limits,
although this reflects an error in the 2MASS J-band magnitude for the former star.
On the other hand, the expected number densities inferred from all three
comparisons are broadly consistent with observational results from the
the subset of the Plane covered by the 2M2nd. 
In particular, almost all of the new proper motion stars identified in the 
LSR survey are either halo subdwarfs or mid-type M dwarfs beyond our
12-parsec limit.

\section {Summary and conclusions}

Based on observations of nearby dwarfs with known trigonometric parallaxes, 
we have developed a set of (J, (J-K$_S$)) and (J-H)/(H-K$_S$) colour-magnitude
criteria designed to isolate ultracool dwarfs with spectral types in
the range M8 to L3, and at distances of less than 12 parsecs from the Sun.
We have applied those criteria to the Second Incremental Release of
the 2MASS database, searching  for candidate nearby dwarfs lying within 10 degrees of 
the Galactic Plane. Working from an initial sample of over 10$^6$ candidates
with $1.0 \le (J-K_S) < 2.1$, these criteria, combined with application of
2MASS confusion flags and elimination of candidate IRAS sources, produce a final
list of 1299 potential nearby dwarfs. Visual inspection of first and second
epoch photographic images form the Digitised Sky Survey shows that all save 2 of those
candidates show no evidence for significant proper motion, and are therefore
likely to be distant red giants or highly reddened earlier-type stars. The
remaining two candidates,  2MASSI 0539248+403844 and 2MASSI 0602305+391059, are
both included in L\'epine 
\etal's (2002) catalogue of new proper motion stars with $0.5 < \mu < 2.0$\pma in
the northern Galactic Plane. 

While these results are sparse statistically, the surface density is consistent
with that derived for ultracool dwarfs of similar spectral types in both
photometric and proper motion surveys covering higher 
galactic latitudes. The implication is that
current estimates of the stellar luminosity function (e.g. PMSU4) are
unlikely to require substantial modification to take account of
a substantial reservoir of undiscovered ultracool dwarfs lying near the
Galactic Plane.

However, our analysis clearly indicates potential means of improving the
efficiency of photometric surveys by an order of magnitude. As discussed in \S3.1, 
despite the significantly increased star density near the Plane, 
the overwhelming majority of the ultracool candidates are visible as isolated
point sources on either the POSS I 103aE or SERC/UKST IIIaF/IVN plates. Matching
data from those surveys directly against the 2MASS database would provide two further
means of refining lists of nearby star candidates:
\begin{itemize}
\item First, red giants and reddened earlier-type stars generally have
bluer optical/near-infrared colours than ultracool dwarfs with comparable (J-K$_S$)
colours. The addition of R-band photometry, even with uncertainties of
$\pm0.25$ magnitudes, should eliminate as many as 90\% of the initial candidate
list. 
\item Second, both of the photographic surveys offer sufficient baseline in time,
relative to 2MASS, that proper motions can be determined to an accuracy of 0.01\pma
or better (equivalent to 1 \kms at 20 parsecs). Given those measurements, the remaining 
candidates can be prioritised in terms of their likelihood of being genuine nearby
dwarfs.
\end{itemize}
The addition of these two techniques will allow us to expand our search not only
to fainter magnitudes and more distant ultracool dwarfs, but also to a 
wider colour range, with correspondingly more complete spectral type coverage\footnote{
An alternative strategy would be to start with an LSR-style proper-motion
selected sample, and then apply photometric criteria to winnow the sample
size to manageable proportions. The effectiveness of this approach rests on
whether digitised image subtraction can be pushed to proper motion limits
of 0.1 \pma or less in crowded fields.}.

As discussed above, these search methods cannot be applied to currently available
databases, since the optical catalogues require detection on blue-sensitive plate material.
The latter requirement is imposed with good reason, since source lists derived
from scans of individual photographic plates include a substantial contribution from
noise sources, generally with derived magnitudes close to the plate limit. 
Cross-referencing against other photographic datasets provides a means of producing 
a more reliable source list. Unfortunately, this practice simultaneously
precludes cross-catalogue comparisons which could identify objects with
extreme colours. This problem can be circumvented by integrating the complete datasets 
from each of these disparate surveys into a single framework 
which provides both access to all sources detected in a single passband, and 
the appropriate software tools for reliable identification of extreme-coloured,
unusual sources. This clearly will be a high priority for the forthcoming  
National Virtual Observatory.

\acknowledgements

Thanks to K. Cruz, J. Gizis and the referee, S. Salim, for useful comments.
This research was supported partially by a grant from the
NASA/NSF NStars initiative, administered by Jet Propulsion Laboratory,
Pasadena, CA. This research has made use of 
data products from the Two Micron All Sky Survey,
which is a joint project of the University of Massachusetts and the Infrared
Processing and Analysis Center/California Institute of Technology, funded by
the National Aeronautics and Space Administration and the National Science
Foundation; of the SIMBAD database, operated at CDS, Strasbourg, France; and
the NASA/IPAC Infrared Science Archive, which is operated by the Jet Propulsion
Laboratory, California Institute of Technology, under contract with the
National Aeronautics and Space Administration. \\
The Digitized Sky Survey was produced at the Space Telescope Science Institute under U.S. 
Government grant NAG W-2166. The images of these surveys are based on photographic data 
obtained using the Oschin Schmidt Telescope on Palomar Mountain and the UK Schmidt Telescope. 
The plates were processed into the present compressed digital form with the permission of 
those institutions. \\
The National Geographic Society - Palomar Observatory Sky Atlas (POSS I) was 
made by the California Institute of Technology with grants from the National
Geographic Society. The Second Palomar Observatory Sky Survey (POSS-II) was made by the California Institute of Technology with funds from the National Science Foundation, the National Aeronautics and Space Administration, the National Geographic Society, the Sloan Foundation, the Samuel Oschin Foundation, and the Eastman Kodak Corporation. The UK Schmidt Telescope was operated by the Royal Observatory Edinburgh, with funding from the UK Science and Engineering Research Council (later the UK Particle Physics and Astronomy Research Council), until 1988 June, and thereafter by the Anglo-Australian Observatory. The blue plates of the southern Sky Atlas and its Equatorial Extension (together known as the SERC-J), as well as the Equatorial Red (ER), the Second Epoch [red] Survey (SES) 
and the Galactic Plane I/SR surveywere all taken with the UK Schmidt

{}

\clearpage

\begin{deluxetable}{rrrrrc}
\tabletypesize{\scriptsize}
\tablecolumns{6}
\tablenum{1}
\tablewidth{0pt}
\tablecaption{Ultracool dwarf candidates with $|b| < 10^o$}
\tablehead{
\colhead{ $\alpha$ (2000)} & \colhead{$\delta$} & \colhead {J} &  \colhead {H} 
&\colhead {K$_S$} & \colhead {Comments }}
\startdata
   5:39:24.75&  40:38:43.9&  11.104$\pm$0.021&  10.424$\pm$0.025&  10.004$\pm$0.026& confirmed M8 dwarf \\
   6:02:30.46&  39:10:59.2&  12.285$\pm$0.025&  11.455$\pm$0.027&  10.860$\pm$0.023& confirmed L1.5 dwarf \\
  16:09:20.92& -60:32:49.6&   8.605$\pm$0.015&   7.758$\pm$0.029&   7.179$\pm$0.020& var? \\
  16:57:25.54& -33:33:33.0&   9.955$\pm$0.021&   9.129$\pm$0.027&   8.555$\pm$0.025& \\
  17:07:33.19& -29:32:17.9&  10.416$\pm$0.028&   9.530$\pm$0.026&   9.035$\pm$0.029& \\
  17:08:05.91& -29:11:35.6&   9.670$\pm$0.024&   8.773$\pm$0.035&   8.058$\pm$0.019& on 1st epoch? \\
  17:12:40.87& -31:55:38.3&   9.979$\pm$0.031&   9.111$\pm$0.023&   8.449$\pm$0.032& \\
  17:12:46.01& -35:07:07.5&  13.310$\pm$0.039&  12.423$\pm$0.044&  11.760$\pm$0.041& \\
  17:15:22.06& -33:00:31.2&  11.814$\pm$0.063&  11.142$\pm$0.023&  10.540$\pm$0.032& \\
  17:19:26.84& -31:19:37.0&  11.269$\pm$0.145&  10.478$\pm$0.024&   9.962$\pm$0.028& \\
  17:21:55.37& -33:59:51.4&  13.006$\pm$0.048&  12.136$\pm$0.077&  11.511$\pm$0.054& \\
  17:24:45.66& -30:55:46.4&  12.325$\pm$0.040&  11.522$\pm$0.060&  10.933$\pm$0.053& \\
  17:24:56.59& -35:58:21.9&  12.572$\pm$0.027&  11.700$\pm$0.027&  11.123$\pm$0.026& \\
  17:25:32.08& -31:30:36.5&  12.625$\pm$0.109&  11.742$\pm$0.031&  11.250$\pm$0.034& \\
  17:26:02.24& -30:29:20.1&  11.666$\pm$0.023&  10.901$\pm$0.039&  10.480$\pm$0.030& merged? \\
  17:26:23.81& -30:44:00.9&  12.689$\pm$0.015&  11.808$\pm$0.038&  11.286$\pm$0.036& \\
  17:26:35.26& -35:36:48.9&  12.634$\pm$0.027&  11.737$\pm$0.027&  11.193$\pm$0.026& \\
  17:28:32.21& -30:55:39.4&  12.346$\pm$0.043&  11.628$\pm$0.093&  11.013$\pm$0.079& \\
  17:29:41.91& -31:22:17.7&  12.616$\pm$0.018&  11.736$\pm$0.046&  11.183$\pm$0.041& \\
  17:30:14.43& -25:40:39.9&  13.557$\pm$0.038&  12.668$\pm$0.063&  11.977$\pm$0.045& close pair\\
  17:31:24.00& -27:02:53.2&  13.410$\pm$0.030&  12.557$\pm$0.037&  11.865$\pm$0.025& \\
  17:31:29.46& -29:25:46.6&  12.778$\pm$0.053&  11.930$\pm$0.037&  11.332$\pm$0.035& \\
  17:31:44.28& -27:31:08.6&  12.423$\pm$0.173&  11.544$\pm$0.090&  11.054$\pm$0.045& \\
  17:31:46.99& -29:48:36.9&  12.439$\pm$0.017&  11.553$\pm$0.063&  11.055$\pm$0.051& \\
  17:31:52.59& -29:27:59.8&  13.158$\pm$0.048&  12.271$\pm$0.058&  11.650$\pm$0.044& \\
  17:31:57.20& -34:24:03.4&  11.694$\pm$0.026&  10.889$\pm$0.038&  10.442$\pm$0.038& \\
  17:33:56.56& -18:28:21.1&   9.602$\pm$0.027&   8.736$\pm$0.048&   8.113$\pm$0.022& var? \\
  17:35:55.62& -25:20:15.7&  12.836$\pm$0.040&  11.985$\pm$0.043&  11.381$\pm$0.035& \\
  17:36:07.40& -26:10:56.1&  12.587$\pm$0.038&  11.692$\pm$0.023&  11.145$\pm$0.028& \\
  17:36:26.86& -32:14:52.9&  11.959$\pm$0.028&  11.121$\pm$0.024&  10.660$\pm$0.031& \\
  17:36:36.70& -28:12:06.3&  13.215$\pm$0.108&  12.326$\pm$0.079&  11.707$\pm$0.048& \\
  17:36:59.86& -40:46:38.4&   9.629$\pm$0.021&   8.778$\pm$0.028&   8.169$\pm$0.027& var?\\
  17:38:04.74& -25:45:01.9&  13.042$\pm$0.032&  12.142$\pm$0.044&  11.563$\pm$0.045& close pair \\
  17:38:42.87& -31:06:08.2&  13.541$\pm$0.025&  12.647$\pm$0.032&  11.974$\pm$0.038& \\
  17:38:51.76& -29:16:04.8&  12.384$\pm$0.030&  11.583$\pm$0.044&  11.078$\pm$0.050& \\
  17:39:50.19& -30:00:40.8&  12.501$\pm$0.023&  11.698$\pm$0.053&  11.091$\pm$0.054& \\
  17:40:40.59& -27:56:34.5&  13.347$\pm$0.048&  12.459$\pm$0.055&  11.752$\pm$0.048& \\
  17:41:11.40& -27:20:00.1&  13.022$\pm$0.104&  12.137$\pm$0.035&  11.569$\pm$0.046& \\
  17:41:30.28& -31:06:43.6&  11.912$\pm$0.031&  11.026$\pm$0.030&  10.493$\pm$0.041& \\
  17:42:27.01& -30:16:33.2&  12.368$\pm$0.030&  11.482$\pm$0.025&  10.988$\pm$0.032& \\
  17:44:07.49& -30:39:24.9&  12.336$\pm$0.024&  11.467$\pm$0.026&  10.957$\pm$0.032& \\
  17:44:23.95& -26:09:19.1&  13.496$\pm$0.073&  12.598$\pm$0.062&  11.913$\pm$0.077& \\
  17:47:02.74& -24:50:09.1&  12.693$\pm$0.097&  11.922$\pm$0.059&  11.332$\pm$0.055& \\
  17:48:18.02& -25:11:47.9&  11.185$\pm$0.016&  10.558$\pm$0.044&  10.181$\pm$0.030& \\
  17:48:24.08& -28:18:05.6&  12.629$\pm$0.021&  11.737$\pm$0.037&  11.255$\pm$0.030& \\
  17:50:09.13& -36:54:59.8&   9.662$\pm$0.026&   8.792$\pm$0.030&   8.256$\pm$0.032& on first epoch?\\
  17:50:25.50& -39:01:18.5&  10.472$\pm$0.035&   9.594$\pm$0.033&   9.093$\pm$0.031& \\
  17:50:41.20& -25:17:29.4&  11.757$\pm$0.025&  10.857$\pm$0.026&  10.229$\pm$0.037& \\
  17:50:54.12& -31:51:14.5&  11.749$\pm$0.106&  11.033$\pm$0.049&  10.521$\pm$0.099& \\
  17:52:26.90& -25:52:50.1&  12.865$\pm$0.036&  11.997$\pm$0.034&  11.452$\pm$0.042& \\
  17:52:44.77& -28:06:06.1&  12.294$\pm$0.031&  11.486$\pm$0.038&  10.904$\pm$0.044& \\
  17:53:35.11& -22:37:36.1&  13.536$\pm$0.037&  12.685$\pm$0.058&  11.971$\pm$0.036& \\
  17:54:06.36& -26:16:22.9&  11.755$\pm$0.027&  10.950$\pm$0.034&  10.398$\pm$0.040& barely visible on 1st epoch\\
  17:54:44.05& -24:04:34.8&  11.991$\pm$0.032&  11.116$\pm$0.024&  10.640$\pm$0.033& \\
  17:55:01.80& -26:21:24.6&  11.954$\pm$0.028&  11.126$\pm$0.027&  10.597$\pm$0.028& \\
  17:56:09.80& -22:09:56.2&  13.352$\pm$0.045&  12.456$\pm$0.023&  11.722$\pm$0.047& close pair\\
  17:57:30.63& -23:18:37.7&  12.904$\pm$0.107&  12.130$\pm$0.098&  11.472$\pm$0.201& \\
  17:57:57.57& -28:04:13.4&  12.983$\pm$0.051&  12.101$\pm$0.074&  11.517$\pm$0.053& \\
  17:58:31.27& -24:55:17.2&  13.032$\pm$0.032&  12.135$\pm$0.038&  11.590$\pm$0.058& \\
  17:59:59.92& -21:18:07.5&  13.378$\pm$0.048&  12.504$\pm$0.231&  11.855$\pm$0.052& close pair \\
  18:00:35.95& -16:56:14.0&  11.943$\pm$0.027&  11.075$\pm$0.006&  10.588$\pm$0.116& \\
  18:00:51.09& -22:06:19.5&  12.357$\pm$0.012&  11.583$\pm$0.113&  10.962$\pm$0.043& on 1st epoch \\
  18:01:27.90& -23:27:09.4&  13.075$\pm$0.027&  12.236$\pm$0.034&  11.616$\pm$0.032& \\
  18:01:30.77& -35:52:11.1&  10.000$\pm$0.023&   9.142$\pm$0.022&   8.523$\pm$0.025& \\
  18:01:50.98& -36:59:32.9&   9.618$\pm$0.023&   8.742$\pm$0.012&   8.181$\pm$0.012& \\
  18:02:29.44& -22:08:55.1&  12.190$\pm$0.024&  11.354$\pm$0.024&  10.849$\pm$0.032& \\
  18:03:59.40& -39:55:12.3&   9.568$\pm$0.029&   8.756$\pm$0.022&   8.137$\pm$0.031& \\
  18:04:38.61& -16:58:18.5&  13.710$\pm$0.055&  12.817$\pm$0.071&  12.093$\pm$0.044& \\
  18:06:29.95& -32:54:20.3&  10.154$\pm$0.024&   9.317$\pm$0.025&   8.846$\pm$0.027& \\
  18:06:58.05& -16:03:51.1&  13.697$\pm$0.051&  12.816$\pm$0.037&  12.101$\pm$0.047& \\
  18:07:13.77& -16:52:24.8&  12.761$\pm$0.067&  11.892$\pm$0.055&  11.321$\pm$0.046& \\
  18:07:25.76& -20:54:14.9&  12.900$\pm$0.034&  12.022$\pm$0.053&  11.493$\pm$0.115& \\
  18:08:15.92& -20:08:06.7&  13.060$\pm$0.046&  12.193$\pm$0.028&  11.483$\pm$0.037& triple \\
  18:08:39.45& -14:18:00.5&  12.585$\pm$0.096&  11.735$\pm$0.040&  11.209$\pm$0.016& \\
  18:08:40.45& -30:50:14.4&   9.560$\pm$0.025&   8.667$\pm$0.065&   8.097$\pm$0.017& \\
  18:09:11.63& -12:30:40.8&  12.182$\pm$0.025&  11.431$\pm$0.043&  10.923$\pm$0.013& \\
  18:10:25.79& -33:22:36.8&  10.522$\pm$0.022&   9.655$\pm$0.024&   9.167$\pm$0.028& \\
  18:11:20.23& -30:23:44.6&   9.962$\pm$0.023&   9.089$\pm$0.021&   8.552$\pm$0.027& \\
  18:12:37.58& -19:05:34.8&  12.206$\pm$0.030&  11.369$\pm$0.028&  10.892$\pm$0.033& \\
  18:12:44.39& -35:41:49.0&   8.913$\pm$0.015&   8.019$\pm$0.051&   7.437$\pm$0.015& \\
  18:13:01.99& -18:08:49.3&  10.904$\pm$0.029&  10.114$\pm$0.030&   9.675$\pm$0.027& \\
  18:13:04.93& -18:08:21.4&  12.115$\pm$0.030&  11.301$\pm$0.029&  10.833$\pm$0.027& \\
  18:13:06.46& -17:53:42.7&  11.710$\pm$0.024&  10.845$\pm$0.021&  10.313$\pm$0.030& \\
  18:13:17.62& -18:06:50.9&  11.787$\pm$0.025&  10.997$\pm$0.021&  10.564$\pm$0.030& \\
  18:13:21.91& -17:51:36.0&  10.731$\pm$0.025&   9.857$\pm$0.023&   9.319$\pm$0.032& \\
  18:14:25.95& -30:35:28.7&  10.553$\pm$0.021&   9.743$\pm$0.035&   9.300$\pm$0.028& \\
  18:18:08.07& -35:05:39.3&   9.710$\pm$0.022&   8.883$\pm$0.043&   8.433$\pm$0.028& \\
  18:18:14.79& -35:22:52.8&   9.768$\pm$0.030&   8.986$\pm$0.029&   8.410$\pm$0.029& \\
  18:20:04.51& -24:21:55.5&   9.529$\pm$0.021&   8.632$\pm$0.023&   8.136$\pm$0.029& \\
  18:20:13.82& -27:09:55.9&  10.018$\pm$0.028&   9.164$\pm$0.022&   8.597$\pm$0.029& \\
  18:21:29.92& -26:19:10.0&   9.826$\pm$0.023&   8.983$\pm$0.031&   8.322$\pm$0.026& on 1st epoch\\
  18:29:13.53& -12:33:26.8&  12.454$\pm$0.036&  11.590$\pm$0.042&  11.057$\pm$0.048& \\
  18:37:57.35&  -3:02:44.5&  12.505$\pm$0.030&  11.723$\pm$0.049&  11.165$\pm$0.082& \\
  18:38:57.38&  -3:42:37.3&  12.027$\pm$0.026&  11.192$\pm$0.022&  10.690$\pm$0.026& \\
  18:39:14.55&  -3:46:16.9&  12.576$\pm$0.027&  11.699$\pm$0.021&  11.153$\pm$0.028& \\
  18:39:30.21&  -5:55:00.2&  12.707$\pm$0.030&  11.830$\pm$0.035&  11.331$\pm$0.037& \\
  18:39:42.51&  -5:56:36.6&  12.503$\pm$0.034&  11.660$\pm$0.043&  11.173$\pm$0.057& \\
  18:42:22.97& -15:10:32.9&   9.829$\pm$0.023&   8.949$\pm$0.052&   8.429$\pm$0.027& M giant\\
  18:46:36.15& -18:25:21.4&  10.037$\pm$0.026&   9.171$\pm$0.031&   8.591$\pm$0.027& \\
  19:41:44.33&  22:09:04.5&  11.908$\pm$0.033&  11.121$\pm$0.060&  10.591$\pm$0.039& close double?\\
  20:09:14.16&  33:14:48.2&  12.197$\pm$0.029&  11.324$\pm$0.026&  10.779$\pm$0.039& POSS I image distorted\\
\enddata
\tablecomments{
}

\end{deluxetable}

%\begin{table}
%\tablenum{1}
%\dummytable\label{table-stars}
%\end{table}

\clearpage

\begin{deluxetable}{rrrrrrrrrrcl}
\rotate
\tabletypesize{\scriptsize}
\tablecolumns{12}
\tablenum{2}
\tablewidth{0pt}
\tablecaption{Ultracool candidates from Luyten's surveys}
\tablehead{
\colhead{NLTT} & \colhead {Name} & \colhead {$\alpha$} & \colhead{$\delta$} &
\colhead{b} & \colhead {m$_r$} & \colhead {J} &
\colhead{(J-H) } & \colhead {(H-K$_S$) } & \colhead {Sp. type } &
\colhead{Sel? } & \colhead {Comments }}
\startdata
     1261&LP 585-86 &   0:24:24.6&  -1:58:19&-64&  19.0&  12.02&   0.94&   0.54&M9.5 & N  & BRI0021 \\
     1292&LP 881-64 &   0:24:44.2& -27:08:24&-85&  14.4&   9.26&   0.73&   0.30&M5.5 & N  & GJ 2005 \\
     1470&LP 349-25 &   0:27:56.0&  22:19:32&-40&  17.0&  10.61&   0.64&   0.41&M8 & Y & Note 1 \\
     1934&LP 645-53 &   0:35:44.1&  -5:41:10&-68&  14.9&  10.72&   0.63&   0.37&M5 & Y & Note 2 \\
     3868&LP 647-13 &   1:09:51.2&  -3:43:26&-66&  17.9&  11.69&   0.77&   0.50&M8 &Y  & Note 2 \\
     7395&LHS 1363  &   2:14:12.6&  -3:57:44&-59&  15.5&  10.47&   0.63&   0.37&M6.5  &Y & Note 2  \\
    13580&LP 775-31 &   4:35:16.5& -16:06:57&-37&  17.4&  10.40&   0.62&   0.44&M8 & Y  &Note 3  \\
    15886&LP 779-41 &   5:58:59.8& -17:28:40&-19&  12.6&  10.30&   0.58&   0.54& &N  &Rejected, Note 4  \\
    15887&LP 779-42 &   5:58:59.8& -17:28:40&-19&  12.1&  10.30&   0.58&   0.54& &N  &Rejected, Note 4  \\
    18549&LP 423-31 &   7:52:23.9&  16:12:15& 21&  16.3&  10.83&   0.64&   0.37&M7 &Y  &  \\
    20475&LP 666-9  &   8:53:36.2&  -3:29:32& 26&  17.9&  11.19&   0.72&   0.50&M8 &Y  &LHS 2065  \\
    23785&G 118-43  &  10:15:06.9&  31:25:11& 56&  12.9&   9.41&   0.63&   0.37&M4 & Y & GJ 3590 \\
    28351&LP 375-26 &  11:43:37.8&  24:41:25& 75&  12.8&  11.44&   0.50&   1.74& &N  &Rejected, Note 5 \\
    30451&G 148-8B  &  12:21:27.1&  30:38:36& 83&  15.0&  10.26&   1.01&   0.40& &N  &Rejected, Note 6  \\
    30894&LHS 6234  &  12:29:09.6&  62:39:38& 55&  14.2&  10.34&   0.56&   0.46& &Y  &  \\
    33702&LP 378-1029 &  13:19:33.5&  24:21:58& 83&  12.5&  11.51&   1.92&   0.06& &N  &Rejected, Note 7  \\
    37621&LP 98-79  &  14:30:37.7&  59:43:25& 53&  18.3&  10.77&   0.65&   0.36&M6.5 &Y  &LHS 2930  \\
    38829&LP 914-54 &  14:56:38.3& -28:09:49& 27&  16.4&   9.96&   0.63&   0.41&M7 & Y & LHS 3003  \\
    41545&G 180-11  &  15:55:31.8&  35:12:03& 50&  12.9&   9.00&   0.71&   0.30&M4.5 & N  &G180-11  \\
    42735&LP 862-26 &  16:25:50.3& -24:00:08& 17&  15.3&  11.94&   0.96&   0.45& &N  & Rejected, Note 9 \\
    43925&G 139-3   &  16:58:25.3&  13:58:11& 31&  13.5&   8.86&   0.57&   0.55&M4 &Y  &GJ 3981  \\
    44823&LP 920-54 &  17:28:24.7& -31:40:41&  1&  11.6&   8.56&   0.78&   0.30& & N &Rejected, Note 10  \\
    44850&LP 920-55 &  17:29:17.4& -30:48:37&  1&  11.9&  11.76&   1.57&   0.68& & N &Rejected, Note 10  \\
    45088&T 36      &  17:36:02.7& -30:17:58&  0&  16.2&   9.36&   1.55&   0.74& & N &Rejected, Note 8  \\
    45440&Ross 134  &  17:47:39.5& -22:56:59&  2&  11.9&   9.75&   1.31&   0.23& & N  &Rejected, Note 11  \\
    45643&LP 921-28 &  17:55:21.0& -30:43:53& -3&  -9.0&  10.76&   1.12&   0.54& & N & Rejected, Note 8 \\
    46471&LP 866-19 &  18:26:43.5& -26:18:31& -7&  13.3&  10.26&   0.83&   0.26& & N & Rejected, Note 8 \\
    48781&LP 514-12 &  20:08:22.0&  15:02:34& -10&  11.6&   9.50&   0.88&   0.17& &N  & G 143-33, Rejected, Note 12 \\
    54407&LP 820-64 &  22:38:33.6& -15:17:59&-57&  12.8&   6.58&   0.61&   0.40&M5.5 & Y & Gl 866 \\
    56700&G 275-42  &  23:23:36.6& -22:32:16&-70&  14.2&  11.04&   0.60&   0.48&M3 & Y &Rejected, Note 13  \\
    58476&-10:6203B &  23:56:21.1&  -9:29:57&-68&   9.6&   7.49&   0.19&   0.82& & N &Rejected, Note 14  \\
\enddata
\tablecomments{
Column1 lists the record number from the SGNLTT catalogue (Salim \& Gould, 2003); Column 2 gives the
name, where we supplement the NLTT with designations from the Lowell survey;\\
Columns 3 and 4 give the 2MASS co-ordinates for equinox 2000, and column 5
lists the Galactic latitude; \\
Column 7 lists the red magnitude from the NLTT or LHS; Columns 8 to 10 give
the 2MASS data; \\
Column 11 lists the spectral type, where available, and Column 12 indicates
whether the (J-H)/(H-K$_S$) colours meet our criteria. \\
Notes on individual stars: \\
1. Catalogued by Gizis \etal (2002); \\
2. Included in Paper III of the present series; \\
3. Identified as a nearby star in Paper III and by Scholz \etal (2002); \\
4. Binary star with separation 5 arcseconds - 2MASS photometry unreliable; \\
5. Binary companion of LP 375-27 (m$_r$=12.0), $\Delta=3''$ - 2MASS photometry unreliable; \\
6. Binary companion of G 148-8 (m$_r$=14.9), $\Delta=4.5''$ - 2MASS photometry unreliable; \\
7. Binary companion of BD+25:2621 (m$_r$=11.0), $\Delta=4''$ - 2MASS photometry unreliable; \\
8. Misidentification - no evidence for motion, probably background giant; \\
9. LP 862-26 is not an M dwarf; \\
10. Inconsistent optical and IR photometry; both stars are merged with red field stars in the 
2MASS scans; \\
11. Not identified as a binary by Luyten, but clearly elongated on both UKST/SERC and 2MASS
images - 2MASS photometry unreliable; \\
12. Merged with a field star in the 2MASS scans, and photometry correspondingly
unreliable. UBV photometry by Ryan (1992) indicates G-type colours. Note that SIMBAD
has separate listings for this star under the LP and Lowell names; \\
13. Binary companion of LP 878-90 (m$_r$=14.2), $\Delta=4''$ - 2MASS photometry unreliable; \\
14. Binary companion of BD-10:6203A (m$_r$=7.8), $\Delta=3''$ - 2MASS photometry unreliable.
}
\end{deluxetable}

%\begin{table}
%\tablenum{2}
%\dummytable\label{tabnltt}
%\end{table}

\clearpage

\begin{deluxetable}{rrrrrrrrrrcl}
\rotate
\tabletypesize{\scriptsize}
\tablecolumns{9}
\tablenum{3}
\tablewidth{0pt}
\tablecaption{2MASS data for LSR proper motion stars}
\tablehead{
\colhead{Name} & \colhead {R} &\colhead{(R-J)} & \colhead {J} & \colhead {H} &
 \colhead {K$_S$ } & \colhead {H$_R$} &
\colhead{Class } & \colhead {Comments }}
\startdata
 LSR0131+5246&  18.4&  2.67& 15.733$\pm$0.093& 15.408$\pm$0.129& 14.913$\pm$0.133&  22.00&  sd  &  \\
 LSR0155+3758&  14.8&  4.32& 10.476$\pm$0.020&  9.860$\pm$0.024&  9.564$\pm$0.033&  18.45& disk & M5.0  \\
 LSR0157+5308&  14.9&  2.88& 12.016$\pm$0.035& 11.506$\pm$0.030& 11.276$\pm$0.027&  18.93& disk/sdM & sdM3.5 \\
 LSR0254+3419&  19.2&  \nodata&  $>17$ &  $>16$ &  $>15$&  23.95&  WD  & Note 1 \\
 LSR0258+5354&  14.9&  1.63& 13.266$\pm$0.022& 12.731$\pm$0.029& 12.585$\pm$0.037&  18.58&  sd  & sdK7 \\
 LSR0316+3132&  15.4&  3.84& 11.560$\pm$0.020& 10.955$\pm$0.027& 10.632$\pm$0.023&  19.80& disk & M5.0 \\
 LSR0346+2456&  18.2&  2.47& 15.728$\pm$0.083& 15.204$\pm$0.132& 14.708$\pm$0.112&  23.70&  WD  & Note 2  \\
 LSR0354+3333&  16.4&  4.60& 11.799$\pm$0.031& 11.270$\pm$0.032& 10.886$\pm$0.022&  21.04& disk & M6.0 \\
 LSR0358+8111&  16.9&  2.57& 14.329$\pm$0.031& 13.815$\pm$0.042& 13.635$\pm$0.056&  20.59&  sd  & sdM1.5 \\
 LSR0403+2616&  19.1&  5.60& 13.503$\pm$0.031& 13.041$\pm$0.036& 12.698$\pm$0.035&  23.22& disk &  \\
 LSR0510+2712&  16.2&  2.53& 13.673$\pm$0.028& 13.078$\pm$0.034& 12.853$\pm$0.033&  20.27&  sd  &  \\
 LSR0510+2713&  15.4&  4.71& 10.688$\pm$0.027&  9.953$\pm$0.030&  9.569$\pm$0.031&  19.50& disk & Note 3 \\
 LSR0519+4213&  16.1&  2.46& 13.638$\pm$0.028& 13.089$\pm$0.033& 12.885$\pm$0.029&  21.46&  sd  & esdM3.5\\
 LSR0520+2159&  16.4&  1.99& 14.406$\pm$0.037& 13.955$\pm$0.042& 13.838$\pm$0.053&  20.74&  sd  &  \\
 LSR0521+3425&  15.8&  3.96& 11.844$\pm$0.029& 11.294$\pm$0.031& 10.984$\pm$0.033&  19.35& disk & M5.0 \\
 LSR0522+3814&  14.5&  1.36& 13.143$\pm$0.030& 12.744$\pm$0.039& 12.622$\pm$0.035&  20.66&  WD  & esdM3.0 \\
 LSR0524+3358&  17.5&  2.67& 14.832$\pm$0.037& 14.360$\pm$0.051& 14.096$\pm$0.070&  21.12&  sd  & sdM1.5 \\
 LSR0527+3009&  16.3&  4.07& 12.234$\pm$0.029& 11.726$\pm$0.050& 11.455$\pm$0.040&  20.33& disk & M5.0 \\
 LSR0532+1354&  19.6&  \nodata&  $>17$ &  $>16$ &  $>15$&  23.41&  WD  & Note 1 \\
 LSR0533+3837&  16.3&  2.45& 13.849$\pm$0.033& 13.344$\pm$0.032& 13.092$\pm$0.045&  20.01&  sd  & sdM2.0 \\
 LSR0534+2820&  17.7&  2.50& 15.203$\pm$0.040& 14.711$\pm$0.054& 14.616$\pm$0.077&  21.63&  sd  &  \\
 LSR0539+4038&  17.0&  5.90& 11.104$\pm$0.021& 10.424$\pm$0.025& 10.004$\pm$0.026&  22.12& disk & M8.0 \\
 LSR0541+3959&  17.1&  0.86& 16.239$\pm$0.090& 17.049$\pm$0.100& 15.346$\pm$0.100&  20.86&  WD  &  \\
 LSR0544+2603&  16.1&  0.28& 15.820$\pm$0.075& 15.240$\pm$0.107& 15.110$\pm$0.134&  22.25&  WD  &  \\
 LSR0549+2329&  17.4&  \nodata&  $>17$ &  $>16$ &  $>15$&  23.10&  WD  & Note 1 \\
 LSR0602+3910&  18.3&  6.01& 12.285$\pm$0.025& 11.455$\pm$0.027& 10.860$\pm$0.023&  21.89& disk & L1.5 \\
 LSR0606+1706&  18.1&  2.62& 15.484$\pm$0.054& 15.038$\pm$0.090& 14.803$\pm$0.109&  21.67&  sd  &  \\
 LSR0609+2319&  16.5&  3.37& 13.132$\pm$0.031& 12.656$\pm$0.035& 12.392$\pm$0.033&  21.71& disk/sdM & sdM5.0 \\
 LSR0618+1614&  14.7&  1.97& 12.727$\pm$0.028& 12.235$\pm$0.031& 11.926$\pm$0.026&  18.75&  sd  & sdM2.0 \\
 LSR0621+1219&  17.3&  2.61& 14.693$\pm$0.042& 14.410$\pm$0.055& 14.337$\pm$0.091&  20.85&  sd  &  \\
 LSR0621+3652&  14.5&  1.91& 12.594$\pm$0.021& 12.100$\pm$0.031& 11.937$\pm$0.027&  19.19&  sd  & esdK7 \\
 LSR0638+3128&  19.5&  4.24& 15.263$\pm$0.048& 14.721$\pm$0.069& 14.556$\pm$0.088&  23.44& disk &  \\
 LSR0646+3212&  17.5&  3.53& 13.968$\pm$0.025& 13.433$\pm$0.030& 13.125$\pm$0.037&  21.27& disk/sdM & M5.5 \\
 LSR0658+4442&  17.4&  \nodata&  $>17$ &  $>16$ &  $>15$&  22.98&  WD  & Note 1 \\
 LSR0702+2154&  14.9&  3.76& 11.144$\pm$0.029& 10.618$\pm$0.037& 10.276$\pm$0.030&  18.97& disk & M5.5 \\
 LSR0721+3714&  16.2&  4.40& 11.802$\pm$0.035& 11.128$\pm$0.048& 10.835$\pm$0.037&  20.04& disk & M5.5e \\
 LSR0723+3806&  19.1&  \nodata&  $>17$ &  $>16$ &  $>15$&  23.70&  WD  & Note 1 \\
 LSR0745+2627&  18.4&  \nodata&  $>17$ &  $>16$ &  $>15$&  23.13&  WD  & Note 1 \\
 LSR0803+1548&  16.1&  1.99& 14.107$\pm$0.034& 13.675$\pm$0.065& 13.435$\pm$0.067&  19.66&  sd  & sdM0.0 \\
 LSR1758+1417&  15.8&  0.85& 14.948$\pm$0.044& 14.639$\pm$0.066& 14.617$\pm$0.078&  20.83&  WD  &  \\
 LSR1809-0219&  13.9&  3.77& 10.126$\pm$0.023&  9.605$\pm$0.028&  9.262$\pm$0.027&  17.42& disk & M4.5 \\
 LSR1809-0247&  15.2&  3.80& 11.399$\pm$0.021& 10.911$\pm$0.029& 10.674$\pm$0.026&  20.21& disk & M5.0 \\
 LSR1835+3259&  16.6&  6.33& 10.273$\pm$0.028&  9.582$\pm$0.052&  9.154$\pm$0.038&  20.97& disk & M8.5, Note 4 \\
 LSR1922+4605&  15.1&  1.49& 13.608$\pm$0.029& 13.057$\pm$0.039& 12.868$\pm$0.036&  18.82&  sd  & sdM0.0 \\
 LSR1928-0200A&  15.2&  3.14& 12.059$\pm$0.024& 11.532$\pm$0.029& 11.295$\pm$0.028&  19.87& disk/sdM & M3.5 \\
 LSR1928-0200B&  18.2& 4.26 & 13.943$\pm$0.023 & 13.497$\pm$0.041 & 13.112$\pm$0.040 & 22.87 & disk & M5.5, Note 5 \\
 LSR1933-0138&  13.5&  2.71& 10.788$\pm$0.021& 10.289$\pm$0.031& 10.055$\pm$0.027&  18.26& disk/sdM & M3.0 \\
 LSR1945+4650A&  16.8&  0.91& 15.886$\pm$0.061& 15.596$\pm$0.112& 15.480$\pm$0.201&  20.73&  WD  &  \\
 LSR1945+4650B&  17.0&  1.13& 15.868$\pm$0.062& 15.524$\pm$0.116& 15.234$\pm$0.165&  20.92&  WD  &  \\
 LSR2000+3057&  15.6&  4.95& 10.648$\pm$0.030& 10.113$\pm$0.034&  9.706$\pm$0.019&  21.23& disk & M5.5 \\
 LSR2009+5659&  14.2&  2.32& 11.883$\pm$0.030& 11.379$\pm$0.036& 11.166$\pm$0.024&  18.78&  sd  & sdM2.0 \\
 LSR2010+3938&  13.3&  0.73& 12.567$\pm$0.027& 12.084$\pm$0.030& 11.842$\pm$0.022&  16.85&  WD?  & (sd)M1.5 \\
 LSR2044+1339&  15.4&  4.25& 11.145$\pm$0.030& 10.547$\pm$0.030& 10.256$\pm$0.030&  18.97& disk &  M5.0\\
 LSR2105+2514&  16.1&  1.63& 14.468$\pm$0.030& 13.732$\pm$0.033& 13.194$\pm$0.041&  19.85&  C dwarf  & Note 6 \\
 LSR2107+3600&  15.9&  3.46& 12.441$\pm$0.027& 11.998$\pm$0.029& 11.717$\pm$0.023&  20.23& disk/sdM & M4.5 \\
 LSR2115+3804&  15.0&  2.05& 12.953$\pm$0.028& 12.469$\pm$0.029& 12.253$\pm$0.025&  18.52&  sd  & esdK7 \\
 LSR2132+4754&  14.5&  3.09& 11.406$\pm$0.027& 10.868$\pm$0.023& 10.674$\pm$0.025&  18.28& disk/sdM & M4.0 \\
 LSR2251+4706&  17.9&  4.40& 13.501$\pm$0.028& 12.944$\pm$0.035& 12.645$\pm$0.030&  21.90& disk & M6.5 \\
 LSR2311+5032&  14.6&  3.80& 10.796$\pm$0.021& 10.227$\pm$0.033&  9.877$\pm$0.024&  18.73& disk & M4.5 \\
 LSR2311+5103&  18.1&  4.85& 13.254$\pm$0.023& 12.684$\pm$0.032& 12.289$\pm$0.026&  21.73& disk & M7.5 \\
 LSR2321+4704&  16.2&  2.25& 13.947$\pm$0.025& 13.400$\pm$0.034& 13.210$\pm$0.034&  20.46&  sd  & esdM2.0s \\
\enddata
\tablecomments{ Spectral types are taken from L\'epine, Rich \& Shara (2003), except
for LSR0510, LSR0539, LSR0602 (this paper) and LSR1835 (Reid \etal, 2003). \\
Notes on individual objects: \\
1. These white dwarfs lie within the area covered by the 2MASS Second Incremental
Data Release, but are not detected. \\
2. For further details, see Hambly \etal (1997). \\
3. LSR0510+2713 - correct position is $\alpha = 5^h 10^m 20^s.1, \delta = +27^o 14' 4''$). \\
4. This M8.5 dwarf lies only 5.67 parsecs from the Sun (Reid \etal, 2003). \\ 
5. LSR1928-0200B - correct position is $\alpha = 19^h 28^m 13^s.3, \delta = -2^o 0' 25''$). \\
6. Further details are given in Lowrance \etal (2003). \\
}
\end{deluxetable}

%\begin{table}
%\tablenum{3}
%\dummytable\label{tablsr}
%\end{table}

\clearpage

\centerline {Figure captions}

\figcaption{ The (M$_J$, (J-K$_S$)) main-sequence: we plot 2MASS
data for nearby M, L and T dwarfs with trigonometric parallaxes
accurate to better than 10\% (see Paper I and Dahn \etal, 2002).
The solid line marks the linear colour-magnitude relation we have
adopted to define the 12-parsec boundary for candidate ultracool dwarfs.
Crosses mark K and M dwarfs, L dwarfs are plotted as solid points and T
dwarfs as five-point stars. Note that all four L dwarfs lying over
1 magnitude above the mean relation are known near-equal luminosity
binary systems. Those dwarfs are not included in the fit.}

\figcaption{ The number of 2MASS/IRAS matches as a function of the difference in position,
in arcseconds. The dashed line marks a positional difference of 12 arcseconds; we assume
that 2MASS sources with $\Delta \le 12$ arcseconds are associated with the IRAS source.}

\figcaption{ The  (J, (J-K$_S$)) and (J-H)/(H-K$_S$) diagrams for the surviving late-M/early-L
nearby star candidates. 
Reference data for main-sequence dwarfs (solid squares), L dwarfs (solid
points), T dwarfs (5-point stars), halo subdwarfs (from Leggett \etal, 2000; 
open circles), giants (from Alonso \etal, 1998; solid triangles)
and Galactic AGB stars (from Kerschbaun \etal, 2001; open triangles) 
are plotted in the colour-colour diagram. Disk dwarf photometry is taken from
Leggett (1992) and Dahn \etal (2002).  The solid line shows the effect of 
foreground reddening of A$_V$=3 magnitudes. 
The nearby star candidates are plotted as crosses, with
the dotted lines outlining the final colour-colour selection criteria. }

\figcaption{ The range of spectral types and optical/infrared colours corresponding to the
(J-H)/(H-K$_S$) limits outlined in the text. Crosses mark data for for nearby stars and
brown dwarfs; solid  points identify dwarfs with colours meeting our selection criteria.}

\figcaption{ The ($\alpha$, $\delta$) distribution of the candidate ultracool dwarfs.
Right ascension runs from 0 hours (left) to 24 hours (right). The upper
panel plots all sources; the lower panel plots sources which really lack optical counterparts
on the first epoch plate material.}

\figcaption{ Low resolution spectra of two late-type dwarfs from the LSR survey. 
LSR0539+40 meets the colour-magnitude criteria adopted for the current survey, and
is an M8.5 dwarf at a distance of $\sim10$ parsecs. LSR0510+2713 lies just
outwith our JHK$_S$ two-colour selection criteria, but is also an M8 dwarf at
a comparable distance. A spectrum of the archetypical M8 dwarf, VB 10, is plotted
for comparison.}

\figcaption{ Near-infrared colour-magnitude and colour-colour distributions of stars 
in the Salim/Gould NLTT catalogue which meet our (J, (J-K$_S$)) selection criteria.
The seven dwarfs lying within 10 degrees of the Galactic Plane are plotted
large solid circles the remaining stars are plotted as crosses. The reference
stars in the JHK$_S$ diagram are coded as in Figure 3.}

\figcaption{ Near-infrared data for the 55 stars in the L\'epine \etal\ proper motion
survey with photometry in the 2MASS Second Incremental Data Release.
The solid line in the left panel marks the colour-magnitude selection criteria adopted
in this survey. As discussed in the text, only four LSR stars meet those criteria. 
The errorbars plotted reflect the combined 
photometric uncertainties (from GSC2.1 and 2MASS); the reference
stars in the JHK$_S$ diagram are coded as in Figure 3.}

\figcaption {The reduced proper motion diagram and (R-J)/(J-K$_S$) two-colour
diagram for LSR stars with published 2MASS photometry (solid points). 
The carbon dwarf, LSR2105, is circled. As a reference, we plot
NLTT dwarfs with comparable motions (crosses), while the upper
diagram also includes data for known intermediate- and extreme-metallicity subdwarfs
(solid and open triangles, respectively). The dotted lines provide an
approximate segregation between the different stellar components.  }

\clearpage

\begin{figure}
\figurenum{1}
\plotone{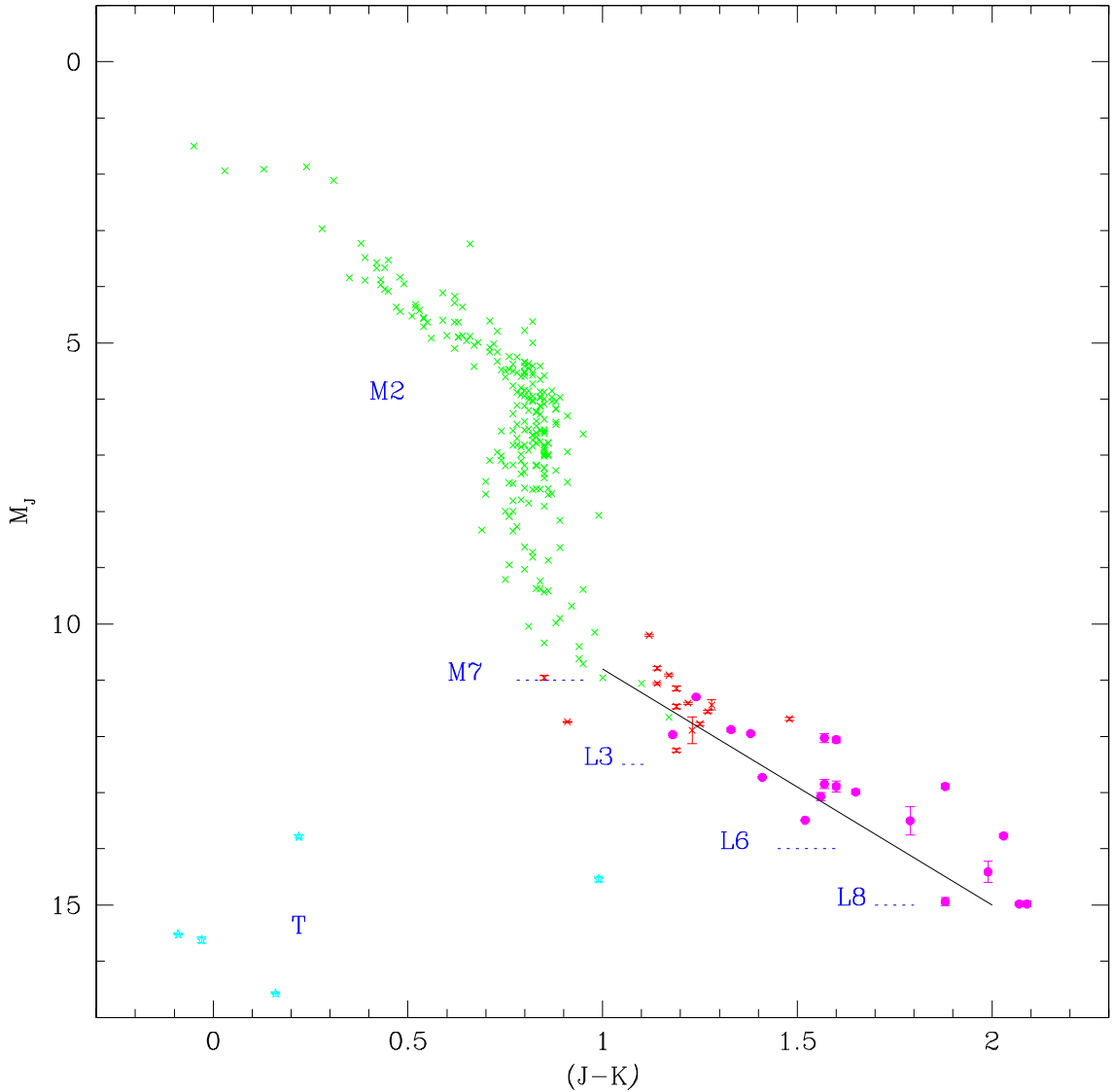}
\caption{The (M$_J$, (J-K$_S$)) main-sequence: we plot 2MASS
data for nearby M, L and T dwarfs with trigonometric parallaxes
accurate to better than 10\% (see Paper I and Dahn \etal, 2002).
The solid line marks the linear colour-magnitude relation we have
adopted to estimate photometric parallaxes for candidate ultracool dwarfs.
Crosses mark K and M dwarfs, L dwarfs are plotted as solid points and T
dwarfs as five-point stars. Note that all four L dwarfs lying over
1 magnitude above the mean relation are known near-equal luminosity
binary systems.}
\end{figure}

\begin{figure}
\figurenum{2}
\plotone{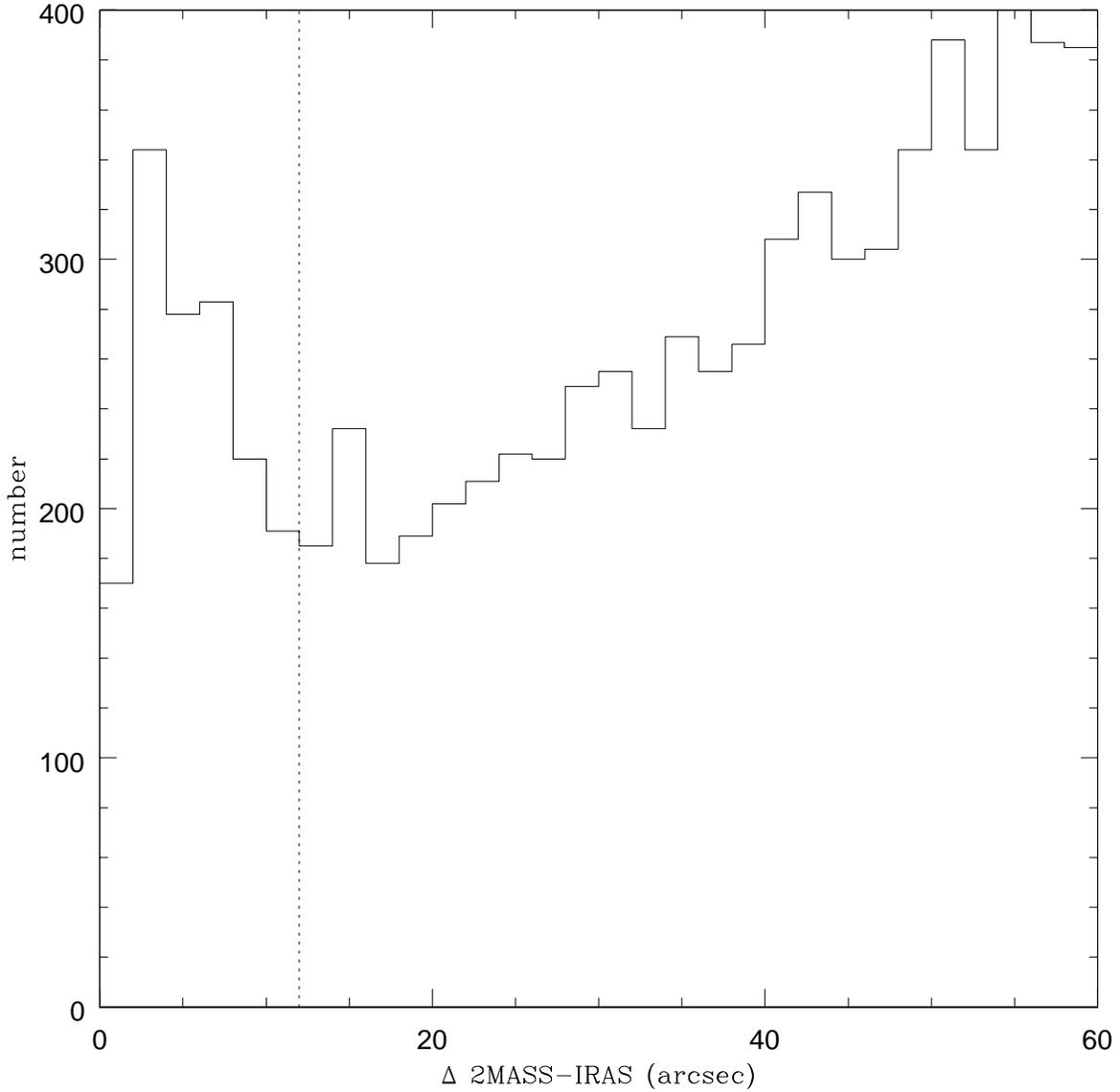}
\caption{ The number of 2MASS/IRAS matches as a function of the difference in position,
in arcseconds. The dashed line marks a positional difference of 12 arcseconds; we assume
that 2MASS sources with $\Delta \le 12$ arcseconds are associated with the IRAS source.}
\end{figure}

\begin{figure}
\figurenum{3}
\plotone{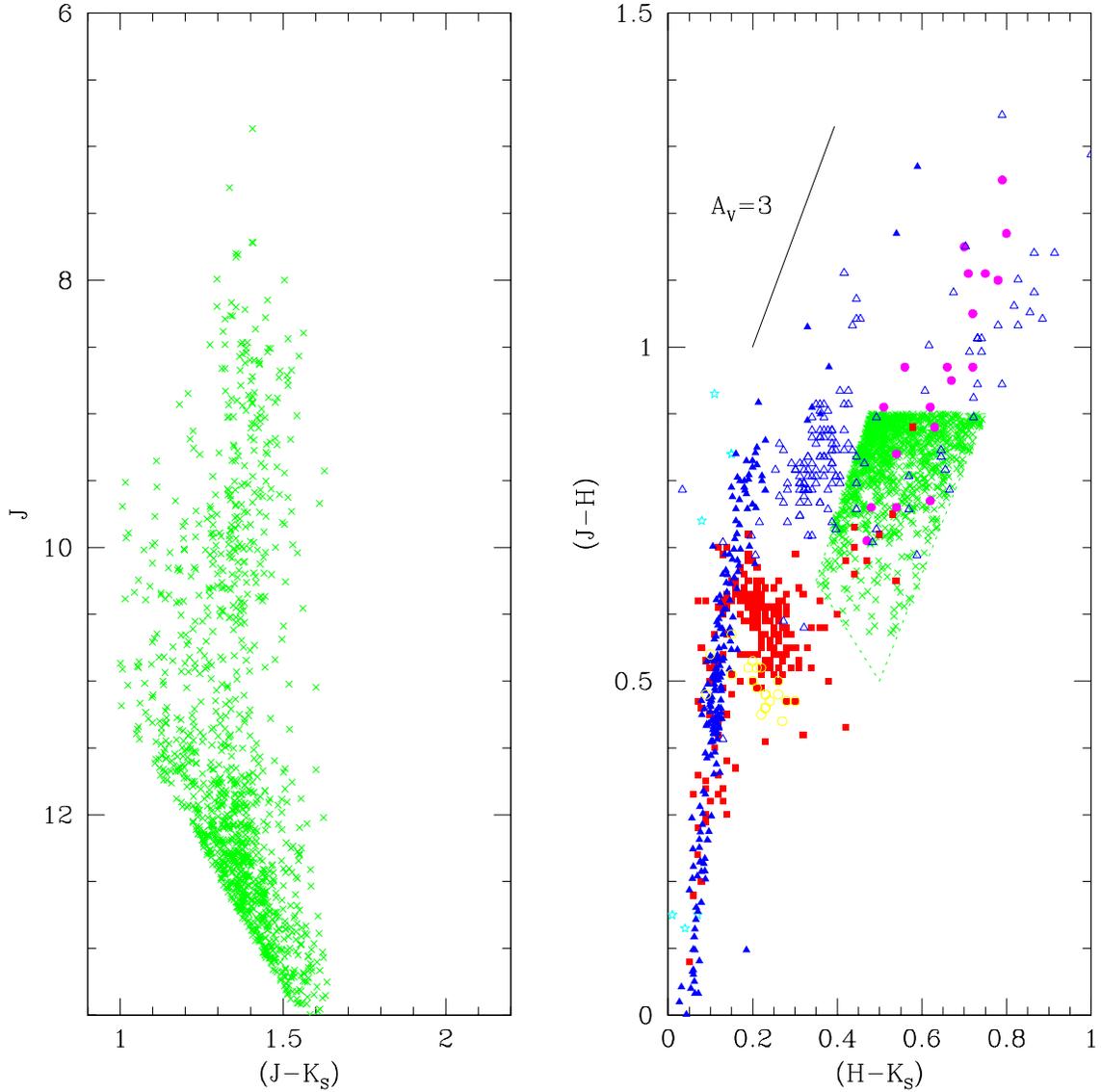}
\caption{ The  (J, (J-K$_S$)) and (J-H)/(H-K$_S$) diagrams for the surviving late-M/early-L
nearby star candidates. 
Reference data for main-sequence dwarfs (solid squares), L dwarfs (solid
points), T dwarfs (5-point stars), halo subdwarfs (from Leggett \etal, 2000; 
open circles), giants (from Alonso \etal, 1998; solid triangles)
and Galactic AGB stars (from Kerschbaun \etal, 2001; open triangles) 
are plotted in the colour-colour diagram. Disk dwarf photometry is taken from
Leggett (1992) and Dahn \etal (2002). The solid line shows the effect of 
foreground reddening of A$_V$=3 magnitudes. 
The nearby star candidates are plotted as crosses, with
the dotted lines outlining the final colour-colour selection criteria. }
\end{figure}

\begin{figure}
\figurenum{4}
\plotone{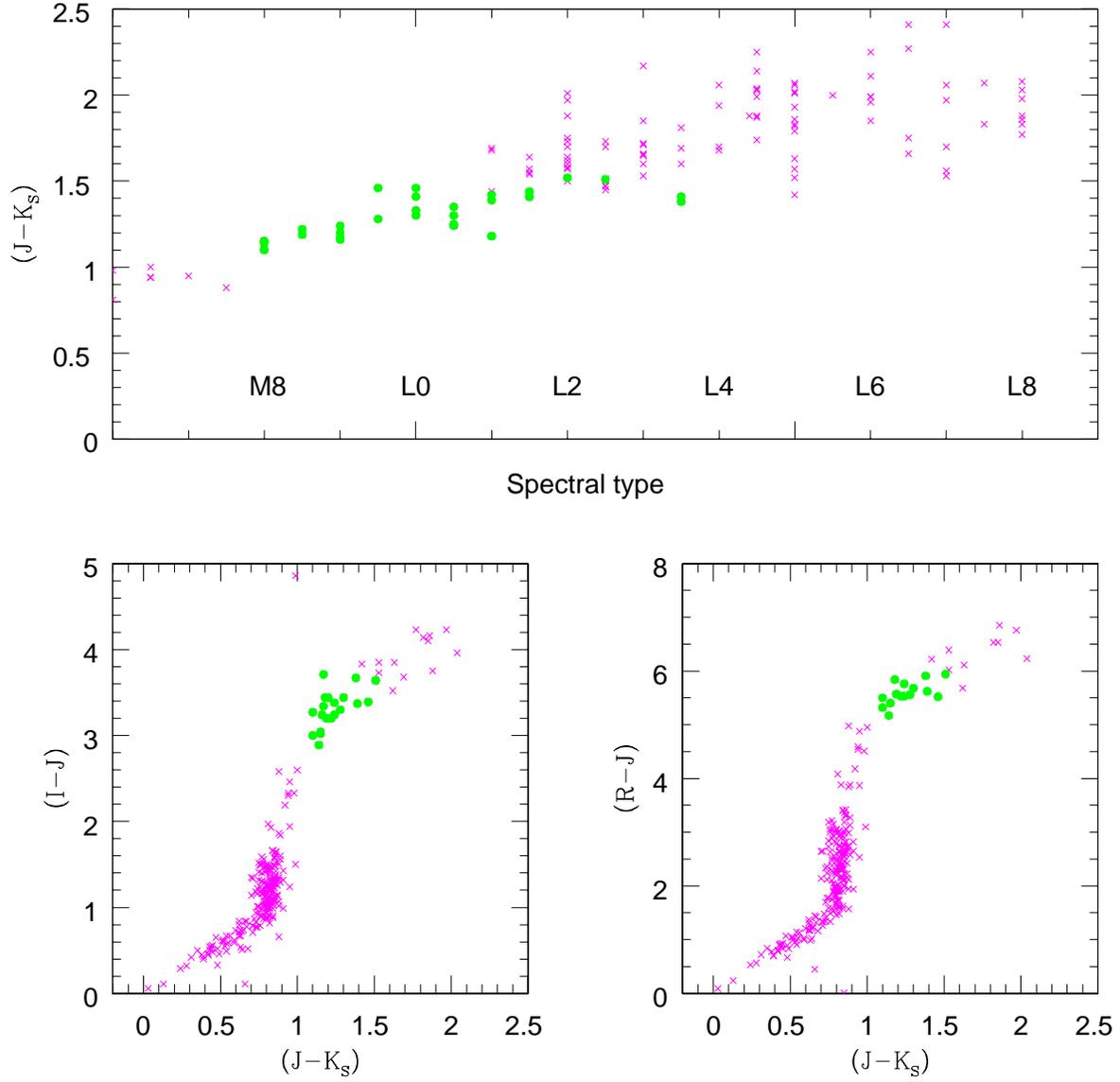}
\caption{ The range of spectral types and optical/infrared colours corresponding to the
(J-H)/(H-K$_S$) limits outlined in the text. Crosses mark data for for nearby stars and
brown dwarfs; solid  points identify dwarfs with colours meeting our selection criteria.}
\end{figure}

\begin{figure}
\figurenum{5}
\plotone{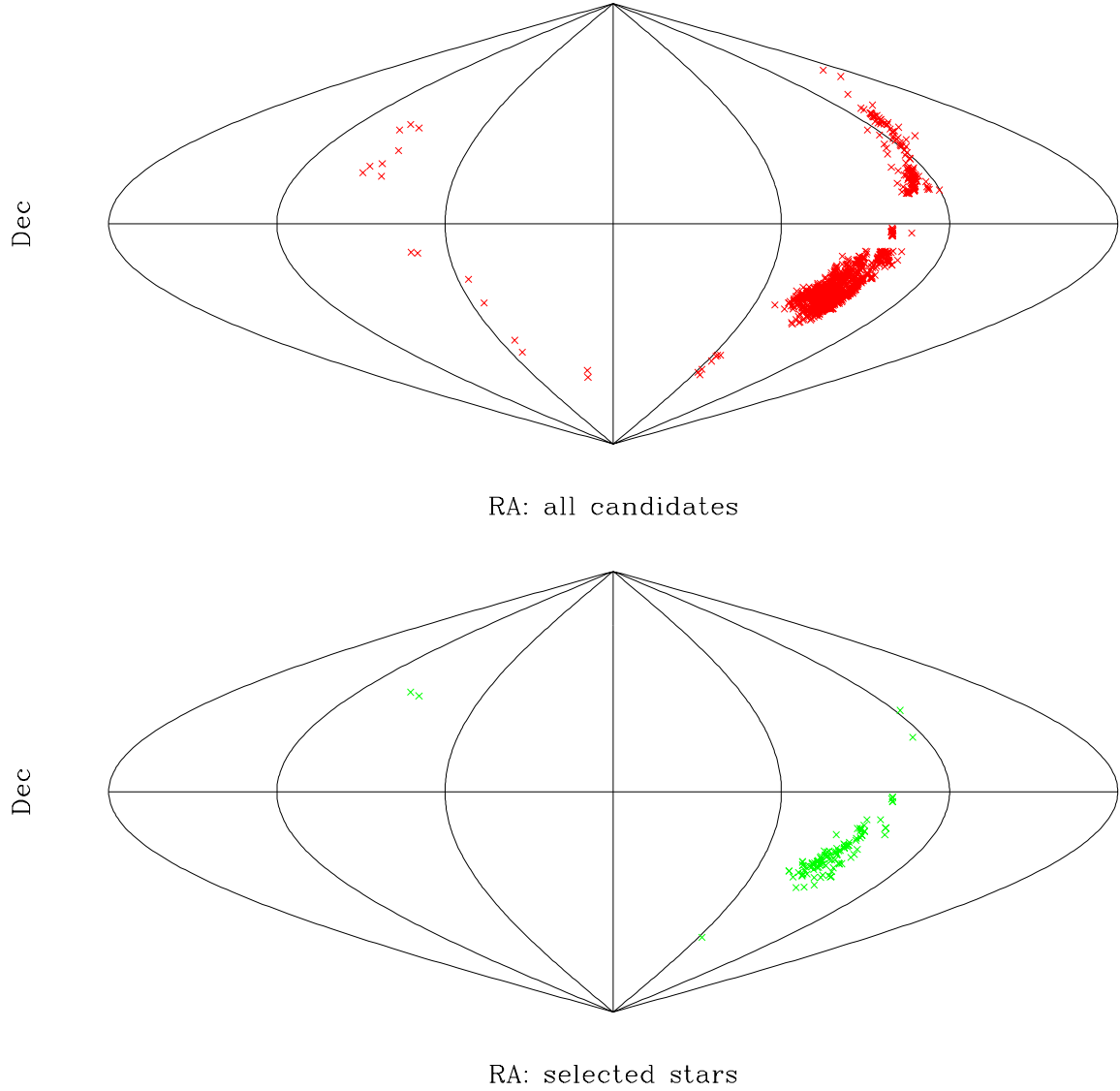}
\caption{ The ($\alpha$, $\delta$) distribution of the candidate ultracool dwarfs. 
Right ascension runs from 0 hours (left) to 24 hours (right). The upper
panel plots all sources; the lower panel plots sources which really lack optical counterparts
on the first epoch plate material.}
\end{figure}

\begin{figure}
\figurenum{6}
\plotone{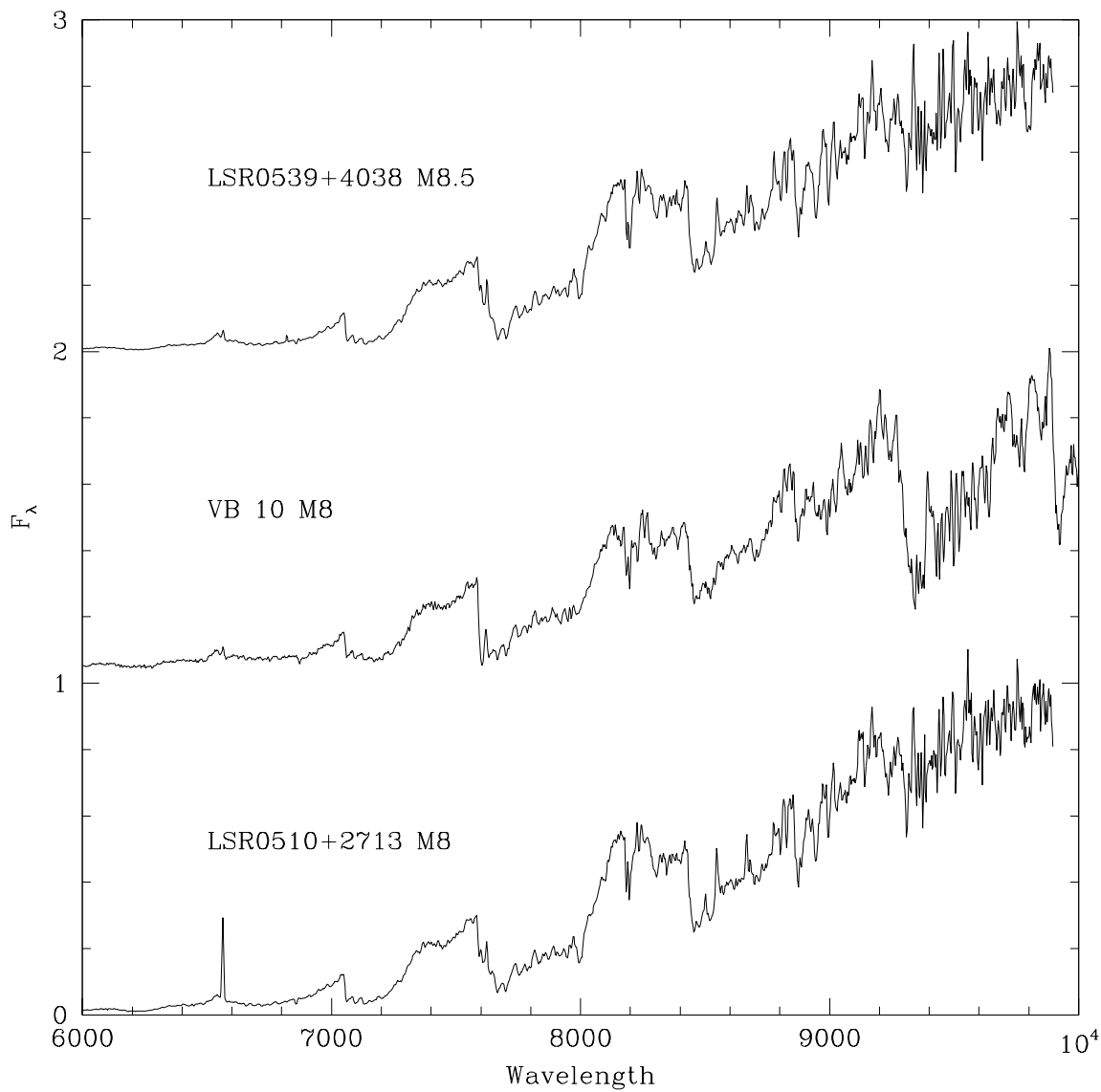}
\caption{ Low resolution spectra of two late-type dwarfs from the LSR survey. 
LSR0539+40 meets the colour-magnitude criteria adopted for the current survey, and
is an M8.5 dwarf at a distance of $\sim10$ parsecs. LSR0510+2713 lies just
outwith our JHK$_S$ two-colour selection criteria, but is also an M8 dwarf at
a comparable distance. A spectrum of the archetypical M8 dwarf, VB 10, is plotted
for comparison.}
\end{figure}

\begin{figure}
\figurenum{7}
\plotone{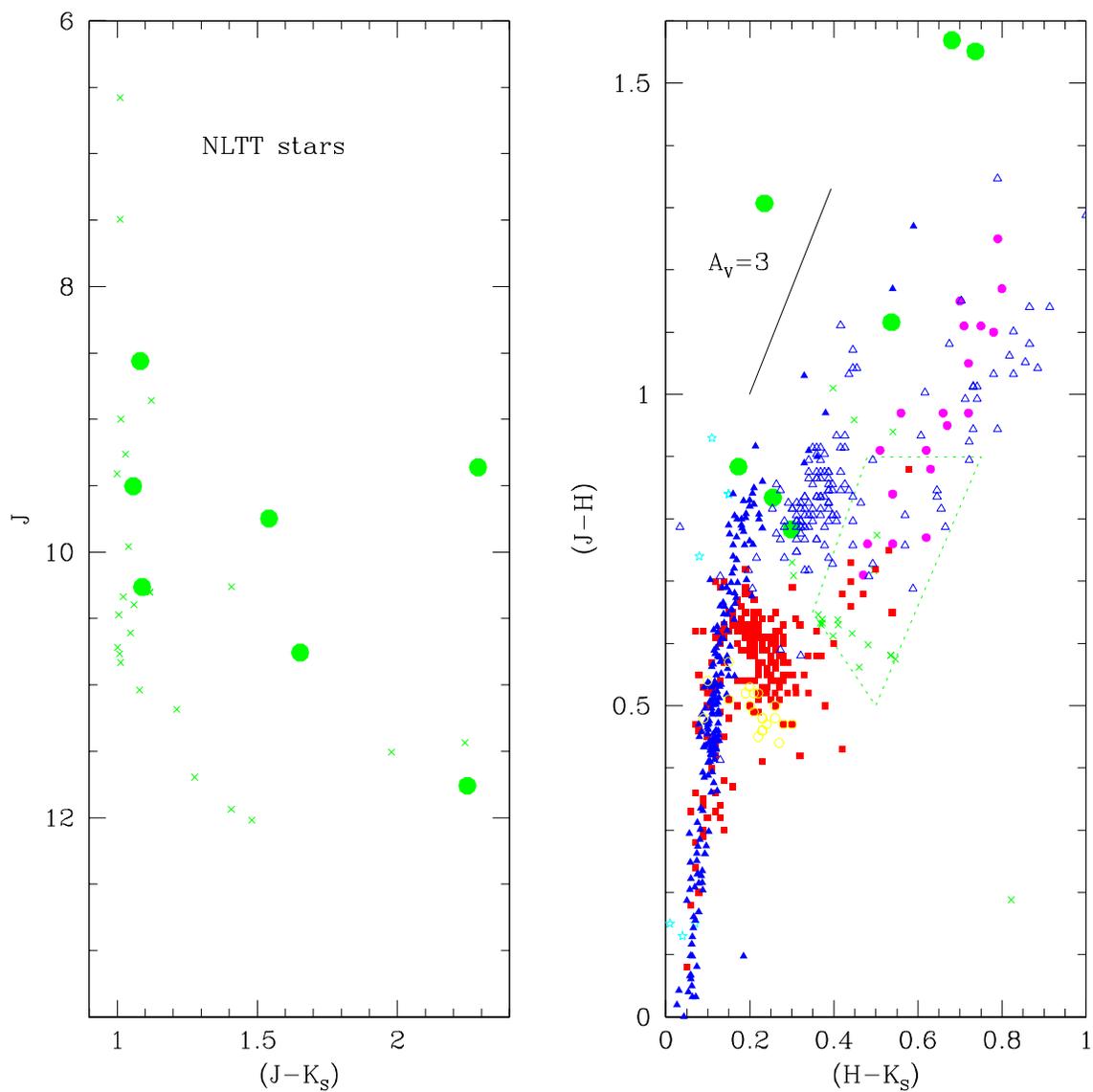}
\caption{ Near-infrared colour-magnitude and colour-colour distributions of stars 
in the Salim/Gould NLTT catalogue which meet our (J, (J-K$_S$)) selection criteria.
The seven dwarfs lying within 10 degrees of the Galactic Plane are plotted
large solid circles the remaining stars are plotted as crosses. The reference
stars in the JHK$_S$ diagram are coded as in Figure 3.}
\end{figure}

\begin{figure}
\figurenum{8}
\plotone{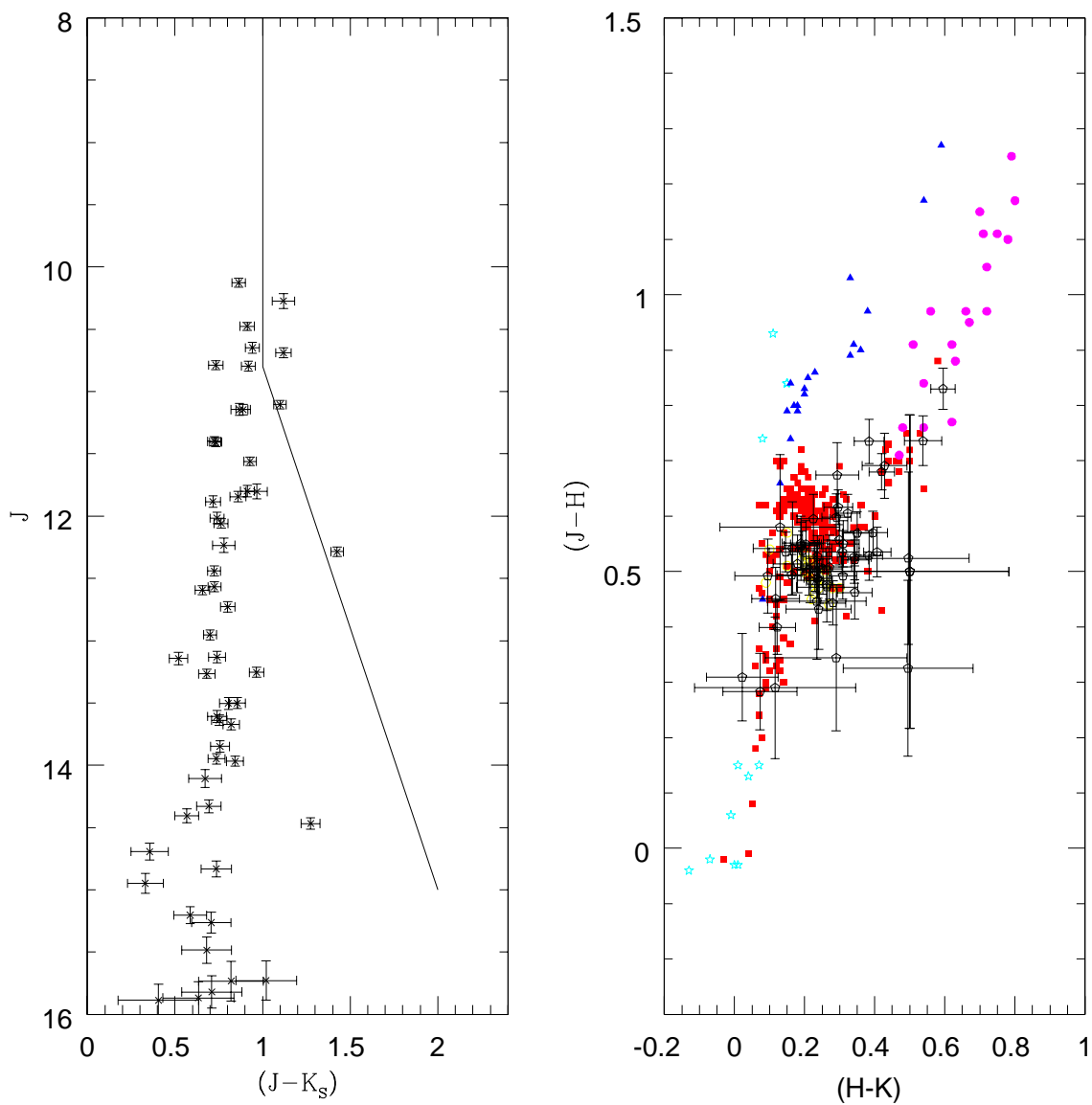}
\caption{ Near-infrared data for the 55 stars in the L\'epine \etal\ proper motion
survey which lie within the region covered by the 2MASS Second Incremental Data Release.
The solid line in the left panel marks the colour-magnitude selection criteria adopted
in this survey. As discussed in the text, only four LSR stars meet those criteria.  
The errorbars plotted reflect the combined 
photometric uncertainties (from GSC2.1 and 2MASS); the reference
stars in the JHK$_S$ diagram are coded as in Figure 3.}
\end{figure}

\begin{figure}
\figurenum{9}
\plotone{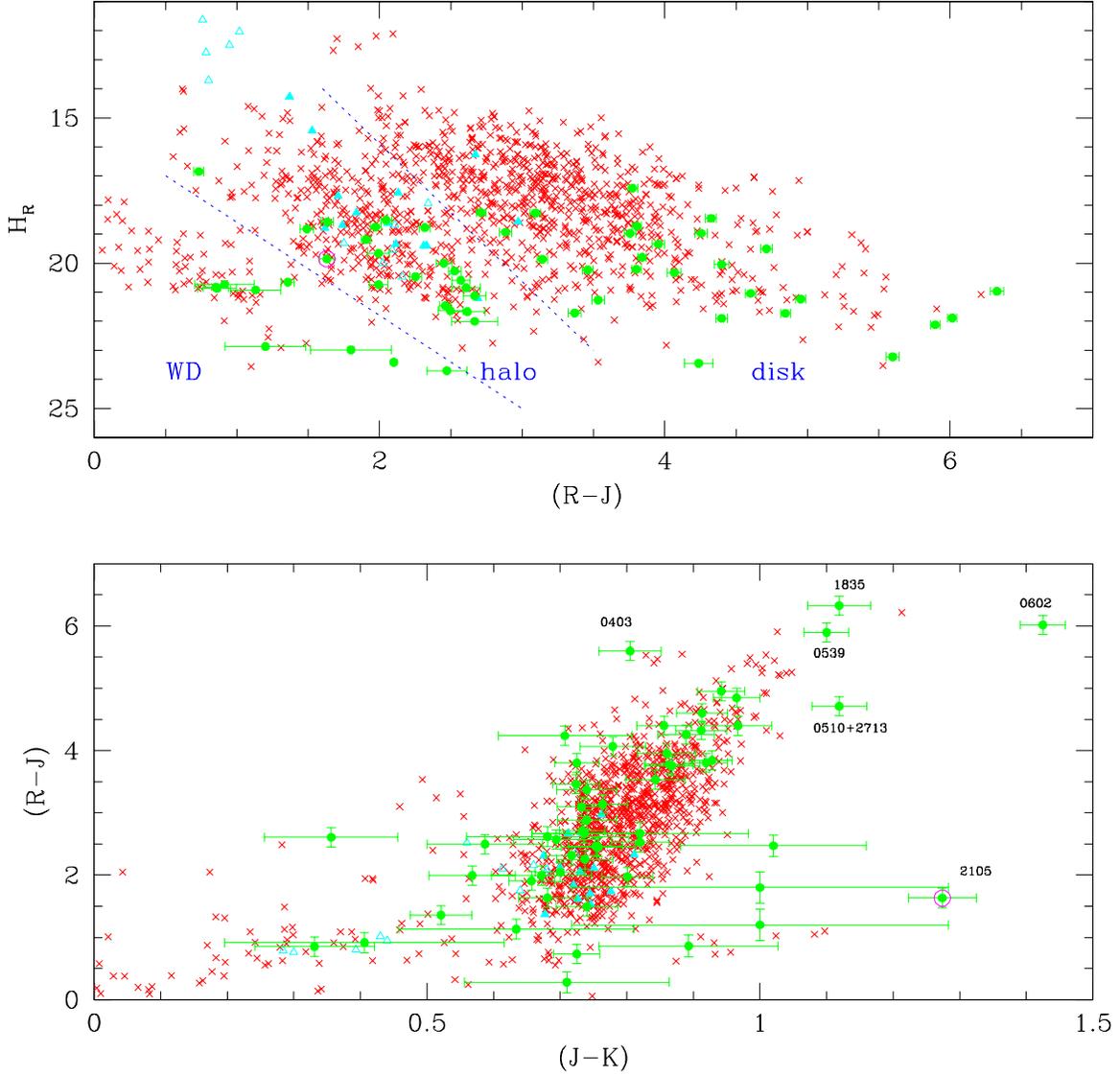}
\caption{ The reduced proper motion diagram and (R-J)/(J-K$_S$) two-colour
diagram for LSR stars with published 2MASS photometry (solid points). 
The carbon dwarf, LSR2105, and the later-type dwarfs are identified in
the two-colour diagram.
As a reference, we plot
NLTT dwarfs with comparable motions (crosses), while the upper
diagram also includes data for known intermediate- and extreme-metallicity subdwarfs
(solid and open triangles, respectively). The dotted lines provide an
approximate segregation between the different stellar components. }
\end{figure}
\end{document}